\documentclass[review,10pt]{elsarticle}
\usepackage{lineno,natbib}
\journal{Computers \& Fluids}

\usepackage{graphicx}
\usepackage{amsfonts}
\usepackage{amssymb}
\usepackage{float}
\usepackage{latexsym}
\usepackage{amsmath}
\usepackage{subfigure}
\usepackage[euler]{textgreek}
\usepackage{textcomp}
\usepackage{hyperref}
\usepackage{float}

\usepackage{algorithm}
\usepackage{algpseudocode}
\usepackage[table]{xcolor}

\DeclareMathOperator\erf{erf}
\DeclareMathOperator\erfc{erfc}

\newcommand{\pd}[2]{\frac{\partial #1}{\partial #2}}
\newcommand{\half}[0]{\frac{1}{2}}
\newcommand{\ub}[0]{\mathbf{u}}
\newcommand{\Db}[0]{\mathbf{D}}

\newcommand{\nb}[0]{\mathbf{n}}
\newcommand{\etal}{\textit{et al.~}}
\newcommand{\ie}{\textit{i.e.}}
\newcommand{\eg}{\textit{e.g.}}
\graphicspath{ {./figures/} }

\begin{document}
	\hypersetup{pdfauthor={Bradley Boyd}}

	\begin{frontmatter}
		
%		\title{A mass-momentum-energy consistent volume-of-fluid approach to simulate interfacial multiphase flows with phase change}
		\title{A consistent volume-of-fluid approach for direct numerical simulation of the aerodynamic breakup of a vaporizing drop}

		\author[add1,add2]{Bradley Boyd\corref{cor1}}
		\ead{bradley.boyd@canterbury.ac.nz}
		\author[add1,add3]{Yue Ling}
		\ead{stanley_ling@sc.edu}
		\cortext[cor1]{Corresponding author. 
			Address:    
			Private Bag 4800, Christchurch 8140, New Zealand
		} 
		\address[add1]{Department of Mechanical Engineering, Baylor University, Waco, Texas 76798, United States}
		\address[add2]{Department of Mechanical Engineering, University of Canterbury, Private Bag 4800, Christchurch 8140, New Zealand}
		\address[add3]{Department of Mechanical Engineering, University of South Carolina, 300 Main St, Columbia, SC 29208, USA}
		%\address[add3]{Department of Mechanical Engineering, Baylor University, Waco, TX 76798, USA}

		 %===============================================================================
		%   Abstract
		%===============================================================================
		
		\begin{abstract}
		A novel simulation framework has been developed in this study for the direct numerical simulation of the aerodynamic breakup of a vaporizing drop. The interfacial multiphase flow  with phase change is resolved using a consistent geometric volume-of-fluid method. The bulk fluids are viscous and incompressible with surface tension at the interface. The newly-developed numerical methods have been implemented in the \emph{Basilisk} solver, in which the adaptive octree/quadtree mesh is used for spatial discretization, allowing flexibility in dynamically refining the mesh in a user-defined region. The simulation framework is extensively validated by a series of benchmark cases, including the 1D Stefan and sucking problems, the growth of a 3D spherical bubble in a superheated liquid, and a 2D film boiling problem. The simulation results agree very well with the exact solution and previous numerical studies. 2D axisymmetric simulations were performed to resolve the vaporization of a moving drop with a low Weber number in a high-temperature free stream. The computed rate of volume loss agrees well with the empirical model of drop evaporation. Finally, the validated solver is used to simulate the aerodynamic breakup of an acetone drop at a high Weber number. A fully 3D simulation is performed and the morphological evolution of the drop is accurately resolved. The rate of vaporization is found to be significantly enhanced due to the drop deformation and breakup. The drop volume decreases nonlinearly in time and at a much higher rate than the empirical correlation for a spherical drop. 

		\end{abstract}

		
	\end{frontmatter}

%\linenumbers

%===============================================================================
%    Introduction
%===============================================================================
\section{Introduction}

%Background of the physical problem

%Review numerical methods to capture sharp interfaces
Direct numerical simulations are crucial to the investigation of a vast array of multiphase flows, such as fuel injection \cite{lefebvre_atomization_2017} and cavitation \cite{boyd_numerical_2019,boyd_numerical_2018, boyd_beamed_2020}, as they are able to provide high-level details that are difficult to measure in experiments \cite{tryggvason_direct_2011}. Accurately resolving the sharp interface separating the gas and liquid phases is critical to interfacial multiphase flow simulations. The methods for numerically resolving the interface are loosely categorized into diffuse-interface and sharp-interface methods. The diffuse-interface methods involve the advection of a volume or mass fraction, which  is allowed to diffuse over a few cells,  and such methods are more popular for compressible flows \cite{saurel_multiphase_1999,johnsen_implementation_2006, boyd_diffuse-interface_2021, boyd_numerical_2021}. For incompressible flows, it is possible to achieve a genuine sharp interface  and to constrain the interface thickness to one cell. Significant progress has been made in the development of interface-capturing methods, including the volume-of-fluid (VOF), front-tracking, and level-set methods, in the past decades \cite{scardovelli_direct_1999, unverdi_front-tracking_1992, sussman_level_1994}. Now it is viable to accurately simulate multiphase flows with complex interface deformation and topology change, assuming the mesh resolution is sufficiently high.

%Review numerical methods to capture sharp interfaces with phase change, including vaporization (both evaporation and boiling) and condensation. 
Multiphase flows with phase change are essential to a wide range of industrial applications, such as the injection of volatile liquid fuels \cite{sher_flash-boiling_2008} and boiling flows \cite{villegas_direct_2017}. The inclusion of phase change introduces additional complexities in resolving interfacial multiphase flows. While a transition from liquid to vapor can happen at temperatures below the saturation temperature, ($T_{sat}$), more rapid vaporization will occur if the phase change happens at the boiling point, $T_{sat}$. In the present study, we only consider phase change due to vaporization as opposed to condensation \cite{son_temperature_2021, samkhaniani_numerical_2016}. More important, we will focus on vaporization that is driven by heat transfer, instead of the slower process of evaporation due to vapor concentration gradients \cite{sazhin_advanced_2006, shao_computational_2018}.  Nevertheless, with minor modifications, the present methods can be used to simulate multiphase flows with condensation and concentration-gradient-driven evaporation.

%Review numerical methods to capture sharp interfaces with phase change
High-resolution detailed numerical simulations have been shown to be an essential tool to investigate liquid-gas multiphase flows with phase change, and various computational methods have been developed to track the sharp interface with interfacial vaporization. Recent numerical approaches in the literature include front tracking \cite{irfan_front_2017}, Lattice Boltzmann \cite{safari_extended_2013, haghani-hassan-abadi_phase-change_2021}, level-set/ghost fluid \cite{gibou_level_2007,lee_direct_2017, shao_computational_2018}, and Volume-of-Fluid (VOF) methods \cite{welch_volume_2000, hardt_evaporation_2008,ma_numerical_2013, sato_sharp-interface_2013, perez-raya_modeling_2016, datta_modeling_2017, ding_volume_2017, wilson_phase-change_2019, wang_vaporization_2019, palmore_volume_2019, scapin_volume--fluid_2020, bures_direct_2021, malan_geometric_2021}. We refer the readers to Refs. \cite{kharangate_review_2017, tryggvason_direct_2005} for a more comprehensive review of the numerical approaches to account for phase change.

In the present study, we focus on developing a numerical framework to simulate interfacial multiphase flows with phase change based on the VOF method. The VOF method is based on the one-fluid approach, for which only one velocity field is used for both phases. Without phase change, the velocities for the liquid and gas on the two sides of an interface are identical, and thus the velocities in the cells in the vicinity of the interface are ready to be used to advect the interface. However, when phase change like vaporization occurs, the velocity is discontinuous across the interface, as vaporization creates the Stefan flow in the gas phase. The interface velocity is more similar to the liquid velocity instead of the gas velocity. In such a case, the ``one-fluid" velocity field cannot be directly used to advect the volume-fraction function. In the recent works of Malan \etal \citep{malan_geometric_2021} and Bure\v{s} \etal \citep{bures_direct_2021},  an additional pressure projection step has been used to compute a separate velocity field, \ie, the divergence-free liquid velocity across the interface, which is then used for the VOF advection. It is computationally costly to solve the additional pressure Poisson equation for this additional velocity in the whole domain. As a result, they have included two immersed boundaries on the gas and liquid sides of the interface and only solve the Poisson equation in a reduced region near the interface. Nevertheless, this inclusion of additional immersed boundaries is also undesirable, especially for parallel codes that use a more complicated mesh such as an adaptive octree mesh.

To accurately compute the vaporization rate it is important to solve the thermal diffusion and the temperature field near the interface. The recent works by Gao \etal \cite{gao_effect_2022} and Zhao \etal \cite{zhao_boiling_2022} have used the embedded boundary method \cite{johansen_cartesian_1998,schwartz_cartesian_2006} for the thermal flux calculation in each phase. The embedded boundary method typically requires the embedded boundary/surface to be continuous at the cell surfaces. However, this is not guaranteed in the Piece-wise Linear Interface Calculation (PLIC) VOF reconstruction. The different intersecting positions between the cell surface and the reconstructed interfaces from the two adjacent cells create additional complexity in calculating the thermal diffusion flux.  Additionally, the temperature gradient near the interface is highly dependent on the normal, which may exhibit small cell-to-cell fluctuations due to the PLIC VOF reconstruction. These two factors are likely the cause of the artificial velocity currents present in \cite{gao_effect_2022}. In the method of Zhao \etal \cite{zhao_boiling_2022}, the VOF advection is handled using a two-velocity approach, similar to Malan \etal \citep{malan_geometric_2021}, whereas, in the method of Gao \etal \cite{gao_effect_2022}, they distributed the vaporization-induced source term to nearby pure gas and liquid cells but not the interfacial cells, following the method by Hardt and Wondra  \cite{hardt_evaporation_2008}. As a result, the velocity at the interfacial cells is not influenced and can be used for the VOF advection. Gao \etal \cite{gao_effect_2022} diffused the source to wide regions in both gas and liquid sides of the interface for numerical stability. Yet, velocity fluctuations were still observed near the interfaces.

%Goal of study.
The goal of the present paper is to develop a simple and robust VOF method to resolve interfacial multiphase flows with vaporization. The long-term goal is to use the newly developed method to fully characterize the breakup dynamics of vaporizing bulk liquids, such as aerobreakup of drops in a high-temperature and high-speed flow through high-fidelity simulation. Extensive experimental and numerical studies have been conducted to understand drop breakup dynamics when phase change is absent \cite{hsiang_drop_1995, theofanous_physics_2008, meng_numerical_2018, jain_secondary_2019, jackiw_aerodynamic_2021}, however, the interaction between drop deformation/breakup and vaporization is not fully understood. In practical spray applications, due to the large geometric scale and the huge number of drops involved, it is inviable to resolve the interface of each individual drop. For those macro-scale simulations, the Euler-Lagrange point-particle simulations are typically used \cite{duke-walker_evaporation_2021, dahal_numerical_2017, gallot-lavallee_large_2021, salman_lagrangian_2004}. Since the drop-scale interfacial and flow physics are not resolved, the mass and energy transfer between the drop and the surrounding gas must be represented by sub-grid models, similar to the drag model for momentum transfer \cite{maxey_equation_1983, balachandar_turbulent_2010}. The conventional models to predict drop vaporization are empirical correlations for spherical drops \cite{renksizbulut_experimental_1983} and they will be invalid for drops with a Weber number close to or larger than the critical Weber number \cite{hsiang_drop_1995}. Models that can accurately predict the vaporization rate of drops undergoing significant deformation and even breakup remain to be established. High-fidelity interface-resolved simulations enabled by the present study will play an essential role in the development of such models.

%What we do in this paper (Novelty)
The present method is implemented in the open-source solver \emph{Basilisk} \citep{popinet_gerris_2003,popinet_accurate_2009, popinet_quadtree-adaptive_2015}. Validation tests and simulations are then performed using the modified \emph{Basilisk} code. The key advantage of \emph{Basilisk} is that it provides an infrastructure for octree/quadtree mesh for spatial discretization, which allows an important capability to adapt the mesh in user-defined regions. This adaptive mesh refinement (AMR) feature is crucial for the efficient simulation of interfacial multiphase flows with vaporization. A VOF model with phase change on a quad-tree mesh has been implemented by Wang and Yang \cite{wang_vaporization_2019} in the \emph{Gerris} code, the predecessor to \emph{Basilisk}. One key difference between the two codes is on the parallelization of the adaptive mesh \cite{popinet_gerris_2003,popinet_quadtree-adaptive_2015}. While \emph{Gerris} uses a domain decomposition, \emph{Basilisk} decomposes the tree directly and thus achieves improved parallel performance for large-scale simulations using a large number of processors and refinement levels. Due to the limitation of the \emph{Gerris} code, the VOF model of Wang and Yang \cite{wang_vaporization_2019} was only tested with 2D simulations of drop vaporization. The two very recent works mentioned above by Gao \etal \cite{gao_effect_2022} and Zhao \etal \cite{zhao_boiling_2022}  have also built a phase-change model on the \emph{Basilisk} solver. The present method distinguishes itself from the previous studies by the following essential features. First, the present method has used a consistent approach to advect VOF (mass), momentum, and energy across the interface. This consistency between mass and momentum advection has been shown to be important for interfacial multiphase flows with large-density contrast \cite{zhang_modeling_2020}. The consistency among mass, momentum, and energy is also shown to be important to get accurate pressure and velocity across the interface for compressible flows \cite{zhang_direct_2021}. Second, we have proposed a novel treatment to handle the vaporization-induced volumetric source, by which the VOF represented interfaces can be advected accurately and the Stefan flow in the gas phase can be rigorously captured. This treatment will not require an additional velocity field and solving an additional Poisson equation, as needed in the previous methods \cite{malan_geometric_2021, bures_direct_2021, zhao_boiling_2022} and thus is more efficient and easy to implement. The present method carries a similar spirit to the method of Hardt and Wondra (HW) \cite{hardt_evaporation_2008}, which was adopted by Gao \etal \cite{gao_effect_2022} and others \cite{wang_vaporization_2019, georgoulas_enhanced_2017}. We also distribute the source to cells near the interface but not at the interfacial cells to preserve the correct velocity at the interfacial cells for VOF advection. In the HW method, a diffusion equation is required and the source is distributed to both gas and liquid cells. In contrast, we account only for the contribution of the volumetric source that is induced by density differences between the vapor and liquid, which is responsible for the introduction of the Stefan flow,  and only distribute the source to a compact layer of pure gas cells near the interface in a mass-conservative way.

%Outline of the paper
The rest of the paper is organized as follows. The physical model will be presented in Section \ref{section:governing_eqs}. The numerical methods are then introduced in Section \ref{section:numerical_methods}. The overall simulation approaches are validated through a series of tests in Section \ref{section:validation}. We first start with the classic 1D Stefan and sucking problems (Sections \ref{section:stefan} and \ref{section:sucking}), then we move on to the 3D bubble growth in superheated liquid (Section \ref{section:bubble}). For these cases,  analytical solutions are available to validate the simulation results. Furthermore, we have simulated more realistic cases such as film boiling (Section \ref{section:film_boil}) and vaporization of a moving drop in a high-temperature environment (Section \ref{section:droplet}). The simulation results were then compared against former numerical and experimental studies. Finally, a large-scale simulation of the aerobreakup of a vaporizing drop has been performed to demonstrate the capability of the present method in resolving complex interfacial multiphase flows with vaporization (Section \ref{3D_droplet}).

%===============================================================================
%    Governing Equations
%===============================================================================
\section{Multiphase models and governing equations}
\label{section:governing_eqs}
In the present study, a simulation framework for liquid-gas two-phase flows with phase change is developed. For the velocity field, the two phases are considered as ``one fluid" and share the same velocity field. The momentum conservation equation for both phases is
\begin{align}
	& \rho \left( \frac{\partial \ub}{\partial t}+ \ub \cdot \nabla \ub \right)= - \nabla p + \nabla \cdot (2\mu \Db) + \rho \mathbf{g}+ \sigma \kappa \delta_{\gamma} \nb_{\gamma}
	\label{eq:momentum}
\end{align}
where $\ub$, $p$, $\mu$, $\rho$, $\sigma$, and $\kappa$ are the velocity, pressure, dynamic viscosity, density, surface tension coefficient, and interfacial curvature, respectively. The interface normal is denoted by $\nb_{\gamma}$, where the subscript $\gamma$ indicates properties associated with the interface. The surface tension is a singular force localized on the sharp interface using the Dirac distribution function $\delta_{\gamma}$. The deformation tensor is defined as $\Db = (\nabla \ub + \nabla \ub^T)/2$. Gravity, which can be easily included, is neglected in the cases considered in the present paper.

The two different phases are distinguished by the color function $c$, which follows the advection equation, 
\begin{align}
	& \frac{\partial c}{\partial t}+ \ub \cdot \nabla c = \frac{-s_{\gamma}}{\rho_l}
	\label{eq:vof_advection}
\end{align}
It is taken that $c=1$ and 0 for liquid and gas phases, respectively. Therefore, the volume-averaged color function in a control volume is equivalent to the liquid volume fraction. The density and viscosity are calculated based on the liquid volume fraction using the arithmetic mean: 
\begin{align}
    \rho  & =c\rho_l+(1-c)\rho_g\, ,\\
    \mu  & =c\mu_l+(1-c)\mu_g\, .
\end{align}
The volumetric source term ($s_{\gamma}$) on the right-hand side is associated with phase change. 
The phase change will also modify the continuity equation as 
 \begin{align}
    & \nabla \cdot \ub= s_{\gamma} \left( \frac{1}{\rho_g}-\frac{1}{\rho_l} \right) \,,
    \label{eq:divergence1}
\end{align}
where the subscripts $l$ and $g$ denote the liquid and gas properties, and the velocity field is divergence-free only in the region away from the interface. As will be discussed in the later section, the projection method is employed and as a result, the pressure Poisson equation will be solved to guarantee the projected velocity satisfies the continuity equation. Furthermore, we will distribute the volumetric source at the interface $s_\gamma$ to adjacent pure gas cells, and the discretized distributed volumetric source is denoted by $\hat{s}$.  

The volumetric source term ($s_{\gamma}$) depends on the rate of vaporization ($j_{\gamma}$) and the interfacial area density ($\phi_{\gamma}$), 
%\phi_{\gamma}=A_{\gamma}/V_c
\begin{align}
	& s_{\gamma} = j_{\gamma} \phi_{\gamma}\, .
	\label{eq:sm}
\end{align} 
The rate of phase change ($j_{\gamma}$) is determined based on the heat fluxes from both sides of the interface, 
\begin{align}
	& j_{\gamma} =\frac{1}{h_{l,g}}\left(k_l (\nabla T_l)_{\gamma} \cdot \nb_{\gamma} - k_g (\nabla T_g)_{\gamma} \cdot \nb_{\gamma}\right)
	\label{eq:j_gamma}
\end{align}
where $T$, $k$, and $h_{l,g}$ are the temperature, thermal conductivity, and latent heat, respectively. Note that the rate of phase change here is governed by the temperature gradient, instead of the vapor concentration gradient \cite{sazhin_advanced_2006, shao_computational_2018}. For the expression given here, $j_\gamma >0 $ for vaporization and $j_\gamma <0 $ for condensation. 

The energy (temperature) equations in the gas and liquid regions are \cite{sato_sharp-interface_2013, bures_direct_2021, malan_geometric_2021},

\begin{align}
	& \rho_g C_{p,g} \left( \pd{T_g}{t} +  \ub \cdot \nabla T_g  \right) = \nabla \cdot (k_g \nabla T_g) \, ,
	\label{eq:temp_gas}\,\\ 
	& \rho_l C_{p,l}\left(\pd{T_l}{t} +   \ub \cdot \nabla T_l \right) = \nabla \cdot (k_l \nabla T_l) \, 
	\label{eq:temp_liq}
\end{align}
with the Dirichlet boundary condition at the vaporizing interface $(T_g)_\gamma=(T_l)_\gamma=T_{sat}$.
The isobaric specific heat for gas and liquid are denoted by $C_{p,g}$ and $C_{p,l}$, respectively. Keeping the two temperature fields makes it easier to apply the temperature boundary condition at the interface. It is worth noting that there are no additional source terms in the energy equations due to phase change since it has been implicitly accounted for by the boundary condition at the interface.

%===============================================================================
%    Numerical Method
%===============================================================================
\section{Numerical methods}
\label{section:numerical_methods}

The governing equations are solved using a finite volume approach based on the projection method. 
The advection of the color function is solved using a geometric VOF method \cite{weymouth_conservative_2010}. 
The advection of momentum near the interface is conducted in a manner consistent with the VOF advection \cite{fuster_all-mach_2018, zhang_modeling_2020}. The surface tension term in the momentum equation is discretized using the balanced-force continuum-surface-force method \cite{francois_balanced-force_2006}. The height-function method is used for curvature calculation \cite{popinet_accurate_2009}. A staggered-in-time discretization of the volume-fraction/density and pressure leads to a formally second-order accurate time discretization. 
The quadtree/octree mesh is used to discretize the 2D/3D spatial domains,  providing important flexibility to dynamically refine the mesh in user-defined regions. The adaptation criterion is based on the wavelet estimate of the discretization errors of the user-defined variables \cite{van_hooft_towards_2018}. Since the present study is focused on the new development for the inclusion of phase change, the above methods will be briefly reviewed below (section \ref{section:vof}, \ref{sec:mom_adv}). The detailed implementation of the methods without phase change can be found in previous studies \cite{popinet_accurate_2009, zhang_modeling_2020}. 

To include vaporization in the numerical model, we need to solve the energy equations for both phases to obtain the temperature near the interface (section \ref{section:energy}). Then based on the temperature gradients on both sides of the interface, the vaporization rate is estimated (section \ref{section:dT}). Furthermore, the interface recession due to the reduction in liquid volume (referred to as interface shifting) needs to be accounted for by modifying the VOF field (section \ref{section:infc_shifting}). Finally, the additional volumetric source due to phase change will be added to the pressure equation to account for the non-zero divergence for the velocity near the interface and the resulting Stefan flow due to the expansion of the dense liquid into a gaseous state (section \ref{section:source}). A new treatment is proposed to handle the volumetric source to guarantee that the velocity at the interface is correctly represented and can be directly used in VOF advection. These procedures will be described below in sequence.

%separating interface advection and tracer advection
\subsection{VOF method}
\label{section:vof}
The advection equation  for the color function, Eq.~\eqref{eq:vof_advection}, is solved using a geometric VOF method. The interface in each computational cell is reconstructed as a planar surface based on the cell-average color function (liquid volume fraction). The interface normal $\nb_\gamma$ is computed based on the Mixed Youngs-Centered (MYC) method \cite{aulisa_interface_2007}. The detailed implementation of the method on an octree mesh was given by Popinet \cite{popinet_accurate_2009}. Equation \eqref{eq:vof_advection} can be rewritten in conservative form as 
\begin{align}
\frac{\partial c}{\partial t} = -  \nabla \cdot \left( c \ub\right ) +   c_c \nabla \cdot \ub + \frac{-s_{\gamma}}{\rho_l}
\label{eq:vof_advection_conserve}
\end{align}
where the first two terms on the right-hand side are for the regular VOF method without phase change, while the last term represents the additional shift of interface due to phase change, the discussion of which will be given later in section \ref{section:infc_shifting}. The VOF advection is conducted in a direction-split manner, taking a 2D example, 
\begin{align}
	 \frac{c_{i,j}^{*}- c_{i,j}^{n}}{\Delta t} & = - \frac{  F_{c,i+1/2,j}-F_{c,i-1/2,j}}{\Delta V}+\left(c_c \pd{u}{x}\right)_{i,j}\, ,
	\label{eq:adv-x_color_func}\\
	 \frac{c_{i,j}^{n+1}- c_{i,j}^{*}}{\Delta t} & = - \frac{  G_{c,i,j+1/2}-G_{c,i,j-1/2}}{\Delta V}+\left(c_c \pd{v}{y}\right)_{i,j}\, ,
	\label{eq:adv-y_color_func}
\end{align}
where $\Delta V$ is the cell volume, and the superscript $c^*$ denotes the auxiliary color-function. 
The value of the color function at the cell center is denoted by $c_c$, which is taken to be $c_c=1$ if ${c}>0.5$ and 0 if ${c}<0.5$. 
It was proved by Weymouth and Yue \cite{weymouth_conservative_2010} that the value of $c_c$ must be kept as a constant for all sweep directions to achieve exact mass conservation. The VOF fluxes in $x$ and $y$ directions are denoted as $F_c$ and $G_c$, respectively. The flux on the right surface in $x$ direction, $F_{c,i+1/2,j}$, is calculated as 
\begin{align}
	F_{c,i+1/2,j} = c_{a} u_{f,i+1/2,j} S \, ,
	\label{eq:vof_flux}
\end{align}
where $u_{f,i+1/2,j}$ is the u-velocity at the cell surface and $S$ is the surface area. The volume fraction of liquid that is advected across the cell surface over $\Delta t$ is $c_a$, which is calculated based on the reconstruction of the interface. The VOF fluxes in the other directions are calculated similarly.

\subsection{VOF-consistent momentum advection}
\label{sec:mom_adv}
It is important to advect the momentum across the interface consistently with the VOF (mass) advection, as shown in previous studies \cite{vaudor_consistent_2017, arrufat_momentum-conserving_2020, zhang_modeling_2020}. To make the advection of momentum and mass consistent, the momentum for the liquid and gas phases are handled separately: 
\begin{align}
	 \frac{(c \rho_l \ub)^a - (c\rho_l \ub)^n}{\Delta t} & = - \nabla \cdot (c\rho_l \ub \ub ) + (c\rho_l \ub)_c \nabla \cdot \ub\,, \\ 
	 \frac{((1-c)\rho_g \ub)^a - ((1-c)\rho_g \ub)^n}{\Delta t} & = - \nabla \cdot ((1-c)\rho_g \ub \ub )+ ((1-c)\rho_g \ub)_c \nabla \cdot \ub\,,
\end{align}
where the superscript $^a$ denotes the auxiliary variables accounting only for the advection term.  The momentum is advected as a tracer associated with VOF advection non-diffusely \cite{lopez-herrera_electrokinetic_2015}. The momentum flux for each phase is computed as the product the VOF flux for the corresponding phase and the momentum per unit volume to be advected \cite{zhang_modeling_2020}. Taking the x-momentum for the liquid phase as an example, 
\begin{align}
	 \frac{(c \rho_l u)_{i,j}^{*}- (c \rho_l u)_{i,j}^{n}}{\Delta t} &= - \frac{  F_{u,i+1/2,j}-F_{u,i-1/2,j}}{\Delta V}+\left((c\rho_l u)_c \pd{u}{x}\right)_{i,j}\, ,
	\label{eq:adv-x_xmom}\\
	 \frac{(c \rho_l u)_{i,j}^{a}- (c \rho_l u)_{i,j}^{*}}{\Delta t} & = - \frac{  G_{u,i,j+1/2}-G_{u,i,j-1/2}}{\Delta V}+\left((c\rho_l u)_c \pd{v}{y}\right)_{i,j}\, ,
	\label{eq:adv-y_xmom}
\end{align}
where $F_u$ and $G_u$ denote the $x$-momentum fluxes for in $x$ and $y$ directions. The x-momentum flux on the right surface of the cell is calculated as 
\begin{align}
	F_{u,i+1/2,j} = (\rho_l u)_a F_{c,i+1/2,j} \, ,
	\label{eq:mom_flux}
\end{align}
 where $(\rho_l u)_a$ is the momentum per unit volume to be advected. The Bell-Collela-Glaz (BCG) second-order upwind scheme \cite{bell_second-order_1989} is used for the reconstruction of $(\rho_l u)_a$ in the upwind cell of the surface where the flux is to be evaluated. The generalized minmod slope limiter is employed to compute the gradient. 
 The liquid momentum $(c\rho_l u)_c$ at the cell center is the analog of $c_c$, and $(c\rho_l u)_c=c \rho_l u$ if ${c}>0.5$ and 0 if ${c}<0.5$.
 After the advection of momentum in all directions, the unified velocity is obtained by
\begin{align}
	 \ub^a_{i,j} = \frac{(c \rho_l \ub)_{i,j}^{a} + ((1-c)\rho_g \ub)_{i,j}^{a}}{(c \rho_l)_{i,j}^{a} + ((1-c)\rho_g)_{i,j}^{a}}\, .
\end{align}

%===============================================================================
%    Energy equation
%===============================================================================
\subsection{Energy equations and consistent energy advection}
\label{section:energy}

The energy equations for both phases, \ie, Eqs.~\eqref{eq:temp_gas}-\eqref{eq:temp_liq}, are solved with the Dirichlet boundary condition at the vaporizing interface $(T_g)_\gamma=(T_l)_\gamma=T_{sat}$. The boundary conditions at the interfaces are invoked by setting $T_g=T_{sat}$ in the cells with $f>0$ and $T_l=T_{sat}$ in cells with $f<1$ \cite{lalanne_numerical_2021}. Note that more sophisticated immersed Dirichlet boundary conditions have been proposed recently \cite{gao_effect_2022, zhao_boiling_2022}. Nevertheless, the simple treatment here seems to be sufficient to yield accurate results. 

The energy fluxes for the advection terms, \eg, $ \ub \cdot \nabla (\rho_l C_{p,l} T_l)$, are computed similarly to the momentum fluxes discussed previously, namely the energy is advected as a tracer associated with the VOF advection. In such a way, the numerical diffusion across the interface can be avoided when we advect energy across the interface. Furthermore, the advection methods for mass, momentum, and energy across the interface are consistent. The consistent advection approach is similar to the methods of Zhang \etal \cite{zhang_direct_2021} for compressible interfacial multiphase flows and thus will make the future extension of the present methods toward compressible flows easier. The energy to be advected in the upwind cell  is computed based on the linear reconstruction of the variable using the Bell–Colella–Glaz scheme and the minmod slope limiter \citep{bell_second-order_1989, popinet_gerris_2003}. 

The time integration of the diffusion terms is treated fully implicitly. {When the temperature for the vapor and liquid are lower than $T_{sat}$, there will be no vaporization, since here we neglect the evaporation effect due to the gradient of vapor concentration. There will be no need to make any adjustments to the interfacial temperature. Nevertheless, as the temperature on either side of the interface increases to be over the saturation temperature, which is the scenario of interest in the present study, vaporization will occur. In such a case, the vapor is assumed to be saturated at the interface, and the temperature at the interface is taken to be fixed at the saturation temperature, $T_{sat}$. The heat sink due to the latent heat of vaporization is implicitly included. The interfacial temperature is then used as the boundary condition for solving the diffusion term in each phase. In all the tests considered in the present study, the interfaces are always saturated. }

A potential  improvement for spatial discretization of the diffusion term is to consider a sub-grid embedded  boundary condition for the interface temperature \cite{malan_geometric_2021, bures_direct_2021, gao_effect_2022, zhao_boiling_2022}. Nevertheless, the present methods already yield very accurate predictions as shown later in  the validation studies (section \ref{section:validation}). {Additionally, previous studies using embedded boundary models for the interface have shown artificial velocity fluctuations near the interface \cite{gao_effect_2022}. This is likely due to the use of the VOF interface normal to compute the interfacial temperature gradient: even in simple interface advection cases, the direction of the interface normal in cells exhibits fluctuations between time steps. The fluctuations in the interface normal will result in fluctuations in the computed interfacial temperature gradient and, eventually, the rate of vaporization. } Therefore, if one wants to take advantage of the sub-grid interfacial location, this challenge must be carefully tackled, and such an extension will be relegated to future works.
%===============================================================================
%    Temperature gradient & vaporization rate
%===============================================================================
\subsection{Calculation of vaporization rate}
% calculation of the temperature gradient
\label{section:dT}
The vaporization rate $j_{\gamma}$ is computed in every interfacial cell ($1>c>0$). The determination of $j_{\gamma}$ (Eq.~\eqref{eq:j_gamma}) requires the temperature gradient on both the liquid and gas sides of the interface: $(\nabla T_l)_{\gamma}$ and $(\nabla T_g)_{\gamma}$.  The calculations for the gas and liquid sides are similar. Here we take the gas temperature gradient as an example to demonstrate the procedures. First of all, it is assumed that the temperature gradient is aligned with the interface normal near the interface, therefore, the projection of the temperature gradient to the interface normal is approximated by its magnitude, \ie, $(\nabla T_g)_{\gamma} \cdot \nb_{\gamma} \approx ||(\nabla T_g)_{\gamma}|| $, where $||\cdot ||$ denotes the magnitude of a vector.  The magnitude of the gas temperature gradient at the interface is then obtained by extrapolation from the neighboring pure gas cells (${c}=0$). 
Considering the interfacial cell $\hat{i}, \hat{j}$ in a 2D $5\times 5$ stencil, in which the temperature gradient is approximated as
\begin{align}
    ||(\nabla T_g)_\gamma|| = ||(\nabla T_g)_{\hat{i},\hat{j}}|| =\sum_{i=\hat{i}-2,j=\hat{j}-2}^{i=\hat{i}+2,j=\hat{j}+2} \hat{w}_{i,j} ||\nabla T_{g}||_{i,j},\,
    \label{eq:temp_grad}
  \end{align}
where the normalized weight $w_{i,j}$ for a cell $i,j$ in the stencil is computed as
\begin{align}
    \hat{w}_{i,j} = \frac{w_{i,j}}{\sum _{i=\hat{i}-2,j=\hat{j}-2}^{i=\hat{i}+2,j=\hat{j}+2}w_{i,j}}\,, 
    \label{eq:weight_tg}
\end{align}
where 
\begin{align}
\,w_{i,j}=
\begin{cases}
	\xi_{i,j} ||\mathbf{d}_{i,j}||^2, 	& \text{if}\ c=0\\
	0, 						& \text{if}\ c>0\, . 
\end{cases}	
\label{eq:weight_tg_1}
\end{align}
In interfacial and liquid cells, $w_{i,j}=0$, and in pure gas cells $w_{i,j}$ is associated with the distance vector from the center of the pure gas cell to the center of the interfacial cell, \ie, $\mathbf{d}_{i,j} = \mathbf{x}_{\hat{i},\hat{j}}-\mathbf{x}_{i,j}$, and its projection to the interfacial normal direction, \ie, $\xi_{i,j}=|\nb_{\gamma} \cdot \mathbf{d}_{i,j}|$, see Fig.~\ref{fig:fig_cell}(c).
The central difference approximation is used to calculate the temperature gradient, we have used only the pure gas cells to avoid computing the temperature gradient across the interface. The large stencil used here is to make sure at least one pure gas cell can be found. 
As a result, the primary contributions to the temperature gradient at the interfacial cell are taken from the pure gas cells where the central difference can be used without the temperature from the interfacial cells, see Fig.~\ref{fig:fig_cell}(c). 
Note that the same procedure applies to the temperature gradient on the liquid side $||(\nabla T_l)_{\gamma}||$.

With the gas and liquid temperature gradients in the interfacial cell, the rate of vaporization for an interfacial cell can be computed according to Eq.~\eqref{eq:j_gamma}. The volumetric source in the interfacial cell due to phase change $s_\gamma$ is then computed using Eq.~\eqref{eq:sm}. Note that the interfacial area density in an interfacial cell is evaluated by $\phi_{\gamma}=A_{\gamma}/\Delta V$, where $A_{\gamma}$ is the area of the VOF reconstructed interface in the cell.

\subsection{Vaporization-induced interface shifting}
\label{section:infc_shifting}
\begin{figure}[tbp]
	\begin{center}
		\includegraphics [width=1\columnwidth]{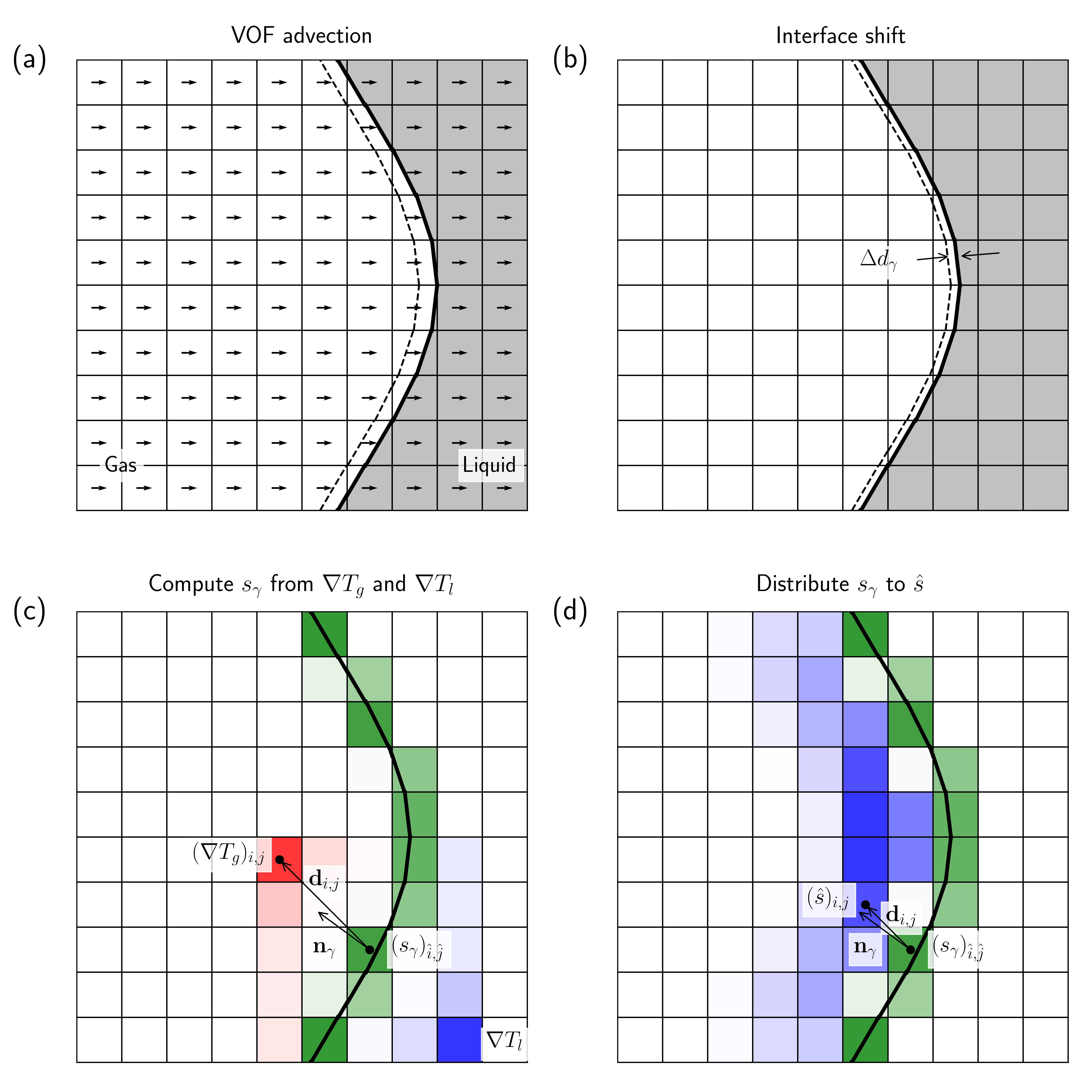}
	\end{center}
	\caption{Key steps of the VOF advection with phase change. (a) the VOF advection based on the liquid velocity, (b) shifting the interface by $\Delta d_{\gamma}$, (c) computing the vaporization source term ($s_{\gamma}$), and (d) distributing the $s_{\gamma}$ to $\hat{s}$ in the near-by pure gas cells.}
	\label{fig:fig_cell}
\end{figure}

The contribution of vaporization on the liquid phase results in an additional shift of the interface toward the liquid side. The vaporization-induced interfacial velocity $\ub_{\gamma}$ is normal to the interface, namely $\ub_\gamma = u_\gamma  \nb_\gamma$, see Fig.~\ref{fig:fig_cell}(b)\cite{malan_geometric_2021}, where $u_\gamma$ is expressed as
\begin{align}
    u_{\gamma} = -\frac{j_{\gamma}}{\rho_l}.\,
    \label{eq:u_shift}
\end{align}
and the negative sign indicates the direction is the opposite of the interface normal. The interface shifting is handled explicitly by shifting the reconstructed VOF interface by the distance over one time step as
\begin{align}
    \Delta d_{\gamma} = -\frac{j_{\gamma}}{\rho_l} \Delta t \,. 
    \label{eq:shift_dist}
\end{align}
The volume fraction of the interfacial cell $c$ is then updated based on the shifted interface. 
Occasionally, the interface may leave the current cell if the liquid volume fraction is very small in the previous time step. In such a case, the interface moves to a neighboring pure liquid cell. Then that liquid cell will become an interfacial cell, with the liquid volume fraction computed based on the interface location. Since phase change induces a new velocity scale $u_\gamma$, it is accounted for in the CFL condition for the time step calculation. 

%%%%%%%%%%%%%% Figure 1 %%%%%%%%%%%%%%%%%%%

\subsection{Pressure equations and volumetric-source distribution}
\label{section:source}
When the higher-density liquid turns into lower-density vapor, the Stefan flow is introduced in the gas phase and the Stefan flow velocity is \begin{align}
    u_\text{st} = j_{\gamma} \left( \frac{1}{\rho_g} - \frac{1}{\rho_l}\right) \,
    \label{eq:u_stef}
\end{align}
 This production of volume due to the different densities is represented by the divergence source term in the continuity Eq.\,\eqref{eq:divergence1}. Since in the projection method the continuity equation is incorporated by solving the pressure Poisson equation, an additional source term appears in the Poisson equation.

The velocity jump across the interface creates a challenge in advecting the color function using the VOF method \cite{malan_geometric_2021, zhao_boiling_2022, gao_effect_2022}. When phase change is absent, the gas and liquid velocities are the same in the interfacial cell, and the velocity for the interfacial cell is ready to be used to compute the VOF fluxes and to advect the interface. However, when the phase change occurs, the gas and liquid velocities jump across the interface. As a result, the velocity in the interfacial cell, which is generally biased towards the gas/vapor velocity, is \emph{not} the correct velocity of the interface. One way to overcome this issue is to solve an additional Poisson equation to extrapolate the liquid velocity across the interface \cite{malan_geometric_2021, bures_direct_2021, gao_effect_2022}. This treatment will result in a divergence-free liquid velocity at the interface for VOF advection and yield accurate results for the vaporizing interface. However, this approach requires an additional Poisson equation, which increases the computational expense. In the method of Malan \etal \cite{malan_geometric_2021}, the boundary conditions on the two artificial embedded boundaries on both sides of the interface are specified to solve the Poisson equation. For moving or deforming interfaces, these embedded boundary conditions need to be imposed every time step, which further increases the computational cost and algorithm complexity, in particular for adaptive octree mesh. 

In the present study, we propose a simple yet accurate method to resolve this numerical challenge in obtaining the current interface advection velocity when  phase change is present. The volumetric source due to vaporization $s_{\gamma}$ is first calculated according to the steps described in Section \ref{section:dT}, based on the temperature gradients for gas and liquid in the interfacial cells. Then instead of applying the volumetric source  $s_\gamma$ right at the interfacial cell, we distribute the volumetric source to the nearest pure gas cells in a $5^3$ stencil in 3D ($5^2$  in 2D).  
The distribution of the volumetric source from the interfacial cell $(\hat{i}, \hat{j})$ to  a pure gas cell $(i, j)$ in the stencil ($\hat{i}-2\le i\le\hat{i}+2$ and $\hat{j}-2\le j\le\hat{j}+2$) is 
\begin{equation}
    (\hat{s}_{i,j})_{\hat{i}, \hat{j}} = \hat{m}_{i,j} (s_\gamma)_{\hat{i}, \hat{j}}\,,
\end{equation} 
where $\hat{m}_{i,j}$ is the normalized weight and is defined similarly to Eq.~\eqref{eq:weight_tg} as
\begin{align}
    \hat{m}_{i,j} = \frac{m_{i,j}}{\sum _{i=\hat{i}-2,j=\hat{j}-2}^{i=\hat{i}+2,j=\hat{j}+2}m_{i,j}}\,,
    \label{eq:weight_vs}
\end{align}
where 
\begin{align}
m_{i,j}=
\begin{cases}
	{\xi_a}/{||\mathbf{d}_{i,j}||}, 	& \text{if}\ c=0\\
	0, 						& \text{if}\ c>0\, .
\end{cases}	
\label{eq:weight_vs_1}
\end{align}
As a result, only pure gas cells in the stencil will receive a distribution. The weight for a gas cell is related to the ratio between the projection of the distance vector $\mathbf{d}_{i,j} = \mathbf{x}_{\hat{i},\hat{j}}-\mathbf{x}_{i,j}$ to the interfacial normal direction, \ie, $\xi_{i,j}=|\nb_{\gamma} \cdot \mathbf{d}_{i,j}|$, and the norm of $\mathbf{d}_{i,j}$. The gas cell in the stencil that is more aligned with the interface normal will thus receive a bigger distribution.
 
Note that the distributed volumetric source in a pure gas cell, $\hat{s}$, is the sum of the source distribution from all the nearby interfacial cells. 
The integration of the distributed source $\hat{s}$ in all pure gas cells is equal to the integration of the volumetric source $s_\gamma$ over all the interfacial cells, 
\begin{align}
	\sum_{1>c_{i,j}>0} (s_{\gamma})_{i,j} = \sum_{c_{i,j}=0} \hat{s}_{i,j}
\end{align}
ensuring the conservation of the generated vapor volume from the whole interface.
Note that here we have used the 2D case to explain the algorithm, and the algorithm is very similar for 3D cases.

Finally, the distributed source is added to the pressure Poisson equation as
\begin{align}
    \nabla \cdot \left( \frac{\Delta t}{\rho} \nabla p \right) = \nabla \cdot \ub^{**} - \hat{s} \left( \frac{1}{\rho_g}-\frac{1}{\rho_l} \right) \,,
    \label{eq:poisson0}
\end{align}
where $\ub^{**}$ is the auxiliary velocity that accounts for all the terms in the momentum equation except the pressure term.

There are several important advantages for the present treatment of the vaporization-induced volumetric source. First of all, by donating the volumetric source from the interfacial to the neighboring gas cells, the velocity in the interfacial cell will not be ``contaminated" by the Stefan flow and will remain as the liquid velocity by which the interface moves (before we impose the additional shifting discussed in section \ref{section:infc_shifting}). Furthermore, the distributed volumetric source in the pure gas cells will induce the Stefan flow in the gas region near the interface. As a numerical approximation, we have moved the volumetric source away from the interface, which will slightly modify the Stefan flow right next to the interface. However, since the relocation distance for $s_\gamma$ to the neighboring cell is typically less than two grid cells, its influence  is very small, as will be shown later in the test results (Section \ref{section:validation}).

The present method also distinguishes itself from the source term distribution method by Hardt and Wondra (HW) \cite{hardt_evaporation_2008, gao_effect_2022} in several important aspects. At first, the present method does not need to solve an additional diffusion equation as required in the HW method. The direct distribution of the source in a compact stencil is beneficial in keeping the effective thickness of the volumetric source small. In general, we have used a $5^3$ stencil centered at the interfacial cell, therefore, the effective thickness is about two cells. It is also possible to use a more compact $3^3$ stencil; however, it will be less robust when there is a large distortion of the interface and there will be no pure gas cells present in the stencil. It is also worth noting that the present method assumes the vapor region is reasonably well resolved and there is at least one pure gas cell in the stencil.
Secondly, in the Hardt and Wondra (HW) method the source is distributed in both pure gas and liquid cells as $s_\gamma/\rho_g$ and $s_\gamma/\rho_l$, respectively, the present method only distributes the volumetric source due to density difference, $s_\gamma (1/\rho_g-1/\rho_l)$, in the pure gas cells and the disappearance of mass in the liquid side is accounted for by shifting the VOF-represented interface explicitly (see section \ref{section:infc_shifting}). While both methods will capture the Stefan flow and interface motion, the present method only needs to modify the velocity field on the gas side and thus has a more compact region of the source distribution. It is worth noting that, vaporization will only modify the velocity and pressure fields and will induce the Stefan flow when the densities of the two phases are different (see Eqs.~\eqref{eq:divergence1} and \eqref{eq:poisson0}). The Stefan flow only appears in the lighter gas phase ($\rho_g\ll \rho_l$), therefore, distributing the volumetric source only in the gas phase is more consistent with the vaporizing flow physics.

\subsection{Staggered-in-time temporal discretization}
The above spatially discrete equations are temporally discretized using the second-order staggered-in-time method in the \emph{Baslisk} solver \cite{popinet_accurate_2009}. The temporally discrete equations are given below, which can be combined with the spatial discretization discussed above to achieve the fully discrete equations. 
\begin{itemize}
	\item Advection equation
\begin{align}
    & \frac{c^{n+\half}-c^{n-\half}}{\Delta t} = - \nabla \cdot \left( c^n \mathbf{u^n}\right ) +  c_c^n \nabla \cdot \mathbf{u^n} + \left(\frac{-s_{\gamma}}{\rho_l} \right)^{n-\half}  \label{eq:vof_advect}
\end{align}
	\item Momentum equation - prediction step
\begin{align}
	 \frac{\big(\rho_l c \ub\big)^{*} - \big(\rho_l c \ub\big)^{n}}{\Delta t} & = -  \ub^n  \cdot \nabla (\rho_l \ub)^n \, , 	\label{eq:mom_adv1}\\ 
	 \frac{\big(\rho_g (1-c) \ub\big)^{*} - \big(\rho_g (1-c) \ub\big)^{n}}{\Delta t} & =- \ub^n  \cdot \nabla (\rho_g \ub)^n\, , 	\label{eq:mom_adv2}\\ 
	 \ub^* & = \frac{\big(\rho_l c \ub\big)^{*}+\big(\rho_g (1-c) \ub \big)^{*}}{\rho_l c^{n+1/2} + \rho_g (1-c^{n+1/2})}\label{eq:mom_adv3}\\
\rho^{n+\half} \frac{\ub^{**} - \ub^* }{\Delta t}  - \nabla \cdot \left( \mu^{n+\half} \Db^{**}\right) & = 
 \nabla \cdot \left( \mu^{n+\half} \Db^*\right)
  + (\sigma \kappa \delta_{\gamma} \nb_{\gamma})^{n+\half}
 \label{eq:u_star}
\end{align}
	\item Pressure Poisson equation
\begin{align}
    & \nabla \cdot \left( \frac{\Delta t}{\rho^{n+\half}} \nabla p^{n+\half} \right) = \nabla \cdot \ub^{**} - \left( \hat{s} \left( \frac{1}{\rho_g}-\frac{1}{\rho_l} \right) \right)^{n+\half} \label{eq:poisson}
\end{align}
	\item Momentum equation - projection step
\begin{align}
 \ub^{n+1}=\ub^{**} - \frac{\Delta t}{\rho^{n+\half}} \nabla p^{n+\half} \label{eq:u_update}
\end{align}
	\item Energy equation    
\begin{align}
	 \frac{(\rho_l C_{p,l} T_l)^{n+\half} - (\rho_l C_{p,l} T_l) ^{n-\half}}{\Delta t}  & =   - \ub^n  \cdot \nabla (\rho_l C_{p,l} T_l  )^{n} + \nabla \cdot (k_l \nabla T_l)^{n+\half}
	 \label{eq:energy_liq}\,
\end{align}
\begin{align}
	\frac{(\rho_g C_{p,g} T_g)^{n+\half} - (\rho_g C_{p,g} T_g)^{n-\half}}{\Delta t}  & = - \ub^n \cdot \nabla  (\rho_g C_{p,g}T_g )^{n} +  \nabla \cdot (k \nabla T_g)^{n+\half}  \label{eq:energy_gas}\,
\end{align}

\end{itemize}

Finally, the overall solution steps are summarized in Algorithm \ref{tab:algor}.
\begin{algorithm}[H]
\caption{Full algorithm summary}\label{alg:cap}
\begin{algorithmic}
\State Initialization of $f$, $\ub$, $T_l$, $T_g$
\While{$t<t_{end}$}
\State Calculate $\Delta t$ based on CFL constraint

\State VOF reconstruction; 
\State VOF advection neglecting the source term due to phase change (Eqs.\ \eqref{eq:adv-x_color_func}, \eqref{eq:adv-y_color_func}, \eqref{eq:vof_advect}); 
\State VOF-consistent advection of momentum and energy (Eqs.~\eqref{eq:mom_adv1}-\eqref{eq:mom_adv3} and Eqs.~\eqref{eq:energy_liq}-\eqref{eq:energy_gas})
\State Solve diffusion terms in energy equations to obtain $T_{l}^{n+\half}$ and $T_{g}^{n+\half}$ (Eqs.~\eqref{eq:energy_liq}-\eqref{eq:energy_gas})
\State Compute the volumetric source term $s_{\gamma}$ (Eqs.~\eqref{eq:sm}-\eqref{eq:j_gamma} and \eqref{eq:temp_grad}-\eqref{eq:weight_tg})
\State Shift reconstructed interface by $\Delta d_{\gamma}$ (Eq.~\eqref{eq:shift_dist}) to account for the phase-change term in the advection equation (Eq.~\eqref{eq:vof_advect})
\State Distribute $s_{\gamma}$ to neighboring pure gas cells and compute $\hat{s}$ 

\State Solve diffusion term in momentum equation to obtain $\ub^{**}$  (Eq. \eqref{eq:u_star})
\State Compute $p^{n+\half}$ by solving the Poisson Eq. \eqref{eq:poisson}
\State Correct $\ub^{n+1}$ by projection (Eq. \eqref{eq:u_update})
\EndWhile
\end{algorithmic}
\label{tab:algor}
\end{algorithm}

\subsection{Quadtree/Octree Mesh}
The physical models and numerical methods described above have been implemented in the \emph{Basilisk} code using adaptive octree/quadtree meshes. The maximum level of refinement $L$ can be compared with a fixed grid resolution of $2^L$ cells in each coordinate direction; \ie, $L9$ corresponds to $2^9=512$ cells in the $x$ direction or $512^3$ cells in 3D. Note that the finite volume cells have equal dimensions (square in 2D or cubic 3D), \ie, $\Delta x_{min}=\Delta y_{min}$. The mesh adaptation algorithm is based on a wavelet-estimated discretization error \cite{popinet_quadtree-adaptive_2015, van_hooft_towards_2018}, where the refinement criteria are based on temperature ($T$), volume fraction ($c$), and velocity ($\ub$). The advantage of using an adaptive mesh is that a higher grid resolution is only used in the user-defined regions; \ie, near the interface, so that the total number of computational cells can be significantly reduced. To maintain the simplicity in the interface temperature gradient calculation (section \ref{section:dT}) and $s_{\gamma}$ distribution (section \ref{section:source}), the mesh near the interface is always refined to the maximum level (5 cells on either side of the interface). This is achieved by refining the mesh based on a level-set function. The addition to the computational cost for this treatment is minor because the temperature gradient is typically large near the interface and requires the maximum level of grid refinement. 

The \emph{Basilisk} code has different parallelization options. In this study, we have used parallelization based on  tree decomposition and MPI \cite{popinet_quadtree-adaptive_2015}. As a result, the computational domain is split into blocks with irregular shapes, instead of simple boxes as in conventional domain decomposition. This advanced parallelization technique guarantees good performance for large-scale simulations using a large number of processors and refinement levels. When an interfacial cell is located at the block boundary, the procedures for the distribution of the volumetric source (section \ref{section:source}) and the shifting of the interface (section \ref{section:infc_shifting}) involve the modification of variables in the ghost cells of an MPI block. An additional MPI communication is thus required for block boundary with interfacial cells. Nevertheless, the computational cost is small compared to that for regular communications between blocks.

%===============================================================================
%    Validation cases
%===============================================================================
\section{Validation}
\label{section:validation}
To validate the present methods and their implementation in the \emph{Basilisk} code, a series of test cases were performed, see Table~\ref{tab:test_cases}. 
The purpose of each test is also listed. The CFL number is set to 0.2 for all cases. The two new velocities induced by vaporization, \ie, the Stefan flow velocity (Eq.~\eqref{eq:u_stef})
and the vaporization-induced interface shifting velocity $u_\gamma$ (Eq.~\eqref{eq:u_shift}) are also accounted for in the calculation of the time step.

%%%%%%%%%%%%%% Table 1 %%%%%%%%%%%%%%%%%%%
 \begin{table*}[tbp]
 \centering
 \begin{tabular}{c c c } 
     \hline
 Case  & Section & Purpose \\
     \hline
 Stefan Problem & \ref{section:stefan}  & Vaporization due to heated gas \\
 Sucking problem & \ref{section:sucking} & Vaporization due to heated liquid\\
Bubble growth & \ref{section:bubble} & 3D interface motion due to vaporization\\
 Film boiling & \ref{section:film_boil} & Complex deformation of vaporizing interface\\ 
     \hline
 \end{tabular}
 \caption{Summary of the validation cases considered in the present paper.}
 \label{tab:test_cases}
 \end{table*}

%===============================================================================
%   1D Stefan problem
%===============================================================================
\subsection{1D Stefan Problem}
\label{section:stefan}
\begin{figure}[tbp]
	\begin{center}
		\includegraphics [width=0.7\columnwidth]{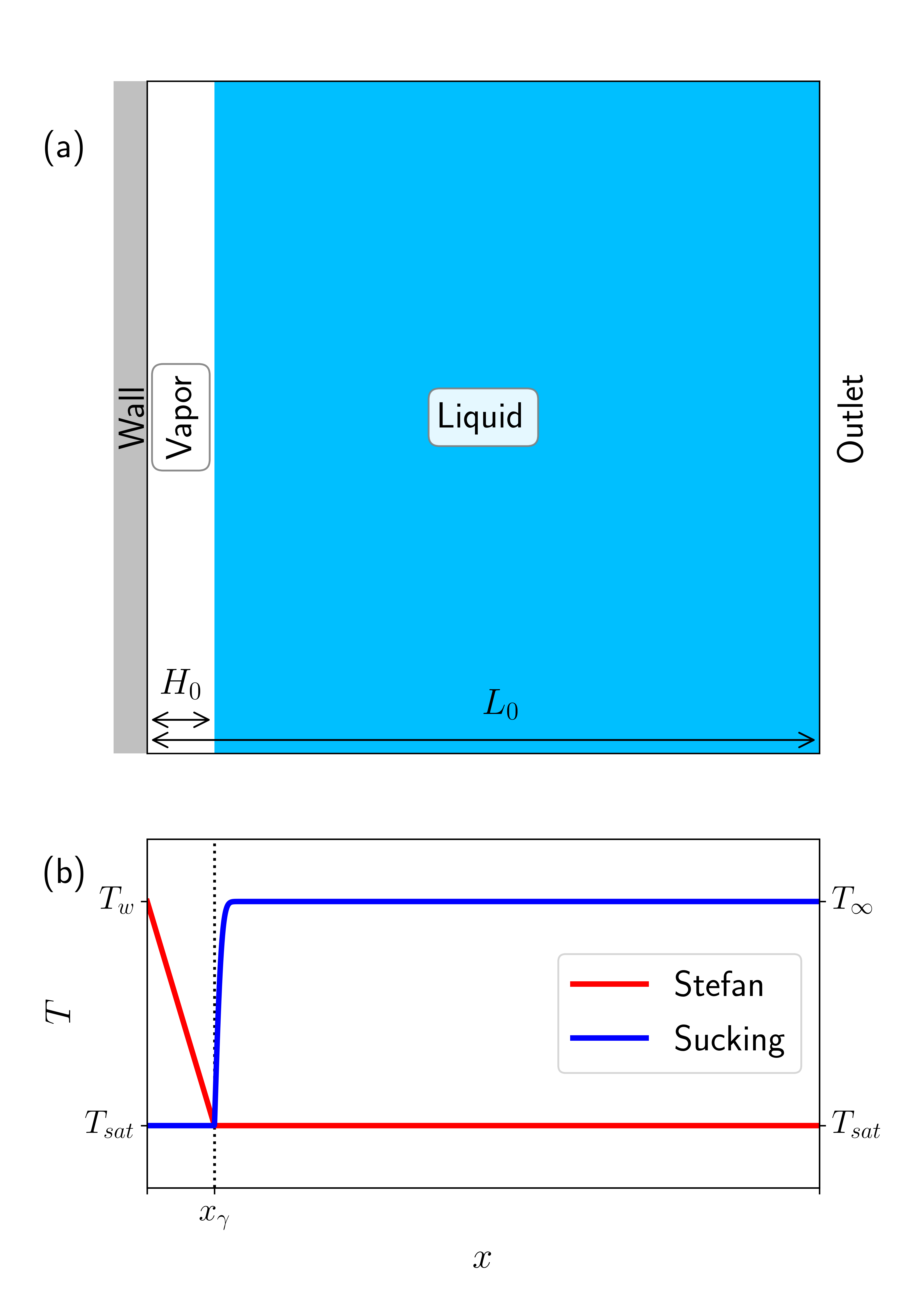}
	\end{center}
	\caption{Schematics for the 1D Stefan and sucking tests: (a) computational domain and (b) temperature distribution. Note that $x_{\gamma}$ is the interface location.}
	\label{fig:stef_suck_schematic} 
\end{figure}

The 1D Stefan problem is a common test case \cite{welch_volume_2000, shao_computational_2018, malan_geometric_2021, sato_sharp-interface_2013, hardt_evaporation_2008, bures_direct_2021, gao_effect_2022}. The domain consists of a vapor region next to a heated wall and a liquid region, see Fig.~\ref{fig:stef_suck_schematic}. The wall  temperature, $T_w> T_{sat}$, is fixed. Initially, the liquid is at saturation temperature ($T_\text{sat}$). 
The liquid is vaporized at the interface due to the heat flux from the vapor. The vapor generated at the interface pushes the liquid to the right. The liquid will move out from the right edge of the domain, on which 
the pressure outlet boundary condition (BC) is imposed. 
The analytical solution for the temporal evolution of the interface position $x_\gamma$ is given as
\begin{align}
	x_{\gamma}(t) = 2 \beta \sqrt{\alpha_g t}, \, 
	\label{eq:stef_interface}
\end{align}
where $\alpha_g=k_g/(\rho_g C_{p,g})$ is the gas thermal diffusivity and $\beta$ is computed from the transcendental equation
\begin{align}
	\beta \exp{(\beta^2)} \erf{(\beta)} = \frac{C_{p,g}(T_w - T_\text{sat})}{h_{lg} \sqrt{\pi}}\,,  
	\label{eq:stef_beta}
\end{align}
where $\erf()$ is the Gauss error function. The temperature in the vapor region, as a function of space and time, is given as
\begin{align}
	T_g(x,t) = T_w + \left(\frac{T_\text{sat} - T_w}{\erf(\beta)}\right)\erf{\left(\frac{x}{2 \sqrt{\alpha_g t}}\right)}.\,
	\label{eq:stef_temperature}
\end{align}

%%%%%%%%%%%%%% Figure 2 %%%%%%%%%%%%%%%%%%%

%%%%%%%%%%%%%% Table 2 %%%%%%%%%%%%%%%%%%%
\begin{table*}[tbp]
 \centering
 \begin{tabular}{l  l l l l l l} 
     \hline
    Property    & \multicolumn{2}{c}{Water} & \multicolumn{2}{c}{Fluid A} & \multicolumn{2}{c}{Fluid B} \\
      &     Liquid & Vapor & Liquid & Vapor & Liquid & Vapor \\
     \hline
 $\rho$ $[kg/m^3]$ & $958.4$ & $0.597$ & $2.5$ & $0.25$ & $200$ & $5$  \\
 $k$ $[W \, m^{-1} \, K^{-1}]$ & $0.679$  & $0.025$ & $0.07$  & $0.007$ & $1$  & $1$\\
 $C_{p,g}$  $[J \, kg^{-1} \, K^{-1}]$  & $4216$ & $2030$  & $2.5$ & $1$ & $200$ & $200$ \\
 %$\mu$ $[Pa \, s]$  & $2.8 \times 10^{-4}$ & $1.26 \times 10^{-5}$& $7.0 \times 10^{-3}$ & $7.0 \times 10^{-4}$  & $0.1$ & $0.005$\\
  $\mu$ $[Pa \, s]$  & $2.8\text{e-}{4}$ & $1.26\text{e-}{5}$& $7\text{e-}{3}$ & $7\text{e-}{4}$  & $0.1$ & $0.005$\\
 %$h_{lg}$  $[J \, kg^{-1}]$ & $2.26 \times 10^6$ &- & $100$ & - & $10^4$ &-\\
 $h_{lg}$  $[J \, kg^{-1}]$ & $2.26\text{e}{6}$ &- & $100$ & - & $1\text{e}{4}$ &-\\
 $T_\text{sat}$  $[K]$ & $373.15$ &- & $1$ & - & $1$ &- \\
 $\sigma$ $[N \, m^{-1}]$ & $0.0728$ &- & $0.001$ & -  & $0.1$ &- \\
     \hline
 \end{tabular}
 \caption{Properties of the saturated water (section \ref{section:stefan}), Fluid A (sections \ref{section:sucking}-\ref{section:bubble}), and Fluid B (\ref{section:film_boil}). }
 \label{tab:properties}
 \end{table*}
 
%%%%%%%%%%%%%% Figure 3 %%%%%%%%%%%%%%%%%%%
\begin{figure}[tbp]
	\begin{center}
		\includegraphics [width=1\columnwidth]{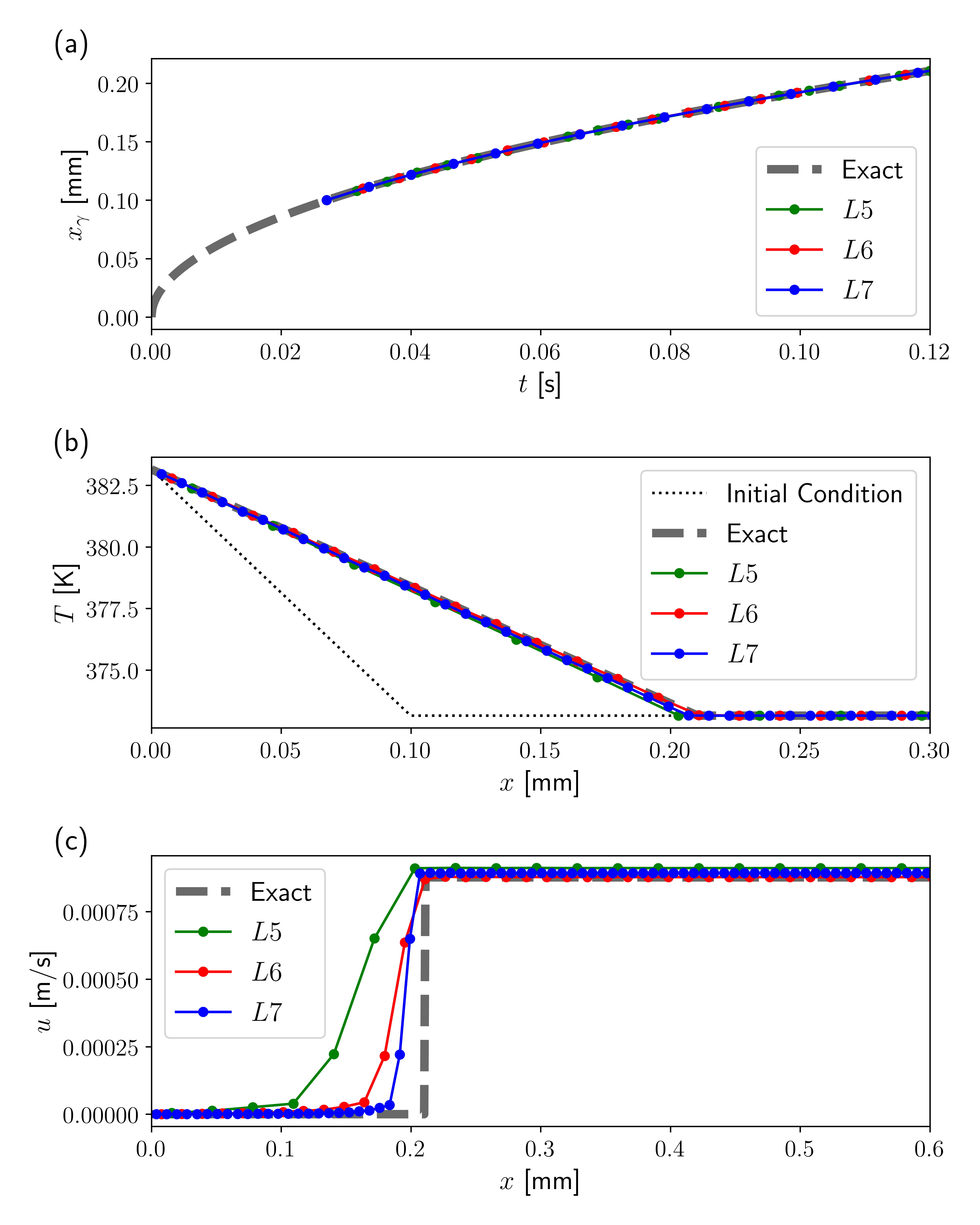}
	\end{center}
	\caption{Numerical results for the 1D Stefan problem for three levels of grid refinement: $L5$, $L6$, and $L7$, compared with the exact solution. (a) Temporal evolution of the interface position $x_\gamma$. (b) The overall temperature $T$ ($T=T_g$ in vapor, $T=T_l$ in liquid, and $T=T_{sat}$ at the interface) and (c) the $u$-velocity distributions at the final time $t=0.12\,$s.}
	\label{fig:stefan_results} 
\end{figure}

Though the flow is 1D, simulations were performed on a 2D square domain with edge length $L_0 = 1\, $mm, see Fig.~\ref{fig:stef_suck_schematic}. The exact solution at $t=0.027$ s was used as the initial condition and the corresponding initial width of the vapor region is $H_0=0.1\,$mm. The fluid is water and the properties are listed in Table~\ref{tab:properties}. The wall temperature is  $T_w=383.15\,$K. The simulation results are presented in Fig.~\ref{fig:stefan_results} for three levels of grid refinement: $L5$, $L6$, and $L7$. It can be observed that the temporal evolution of the interface location for all three meshes matches very well with the analytical solution (see Fig.~\ref{fig:stefan_results}(a)). The simulations have been run to $t=1.2\,$s. The spatial distribution of temperature at the final time is shown in Fig.~\ref{fig:stefan_results}(b). Though we have solved $T_g$ and $T_l$ separately, here we plot the overall temperature $T$, which is defined as $T=T_g$ in vapor, $T=T_l$ in liquid, and $T=T_{sat}$ at the interface. It is clearly seen that the numerical results converge to the exact solution as the mesh is refined. It is also observed that the spatial variation of temperature in the vapor region is approximately linear. That is why even a coarse mesh is sufficient to yield a good approximation. Finally, the spatial variation of the $u$-velocity is shown in Fig.~\ref{fig:stefan_results}(c), compared with the exact solution. Ideally, the velocity in the vapor is zero and that in the liquid is constant in space, $u_l(t) = \dot{x}_{\gamma}$.
Due to the distribution of the volumetric source to the pure gas cells, the numerical results show non-zero vapor velocity in a narrow region on the left of the interface. When the mesh is refined, the thickness of the non-zero velocity region reduces. It is worth noting that the liquid velocity is preserved to be constant on the right of the interface, which converges to the exact value as the cell size decreases. If the source is distributed to both the gas and liquid cells as in the HW method \cite{gao_effect_2022}, numerical smearing will also appear in the liquid velocity.

%===============================================================================
%   1D Sucking problem
%===============================================================================
\subsection{1D sucking problem}
\label{section:sucking}
Another common validation case is the 1D sucking problem, also known as the boiling interface problem  \citep{welch_volume_2000, shao_computational_2018, sato_sharp-interface_2013, bures_direct_2021, zhao_boiling_2022}. The temperature distribution is what differentiates this problem from the Stefan problem, see Fig.~\ref{fig:stef_suck_schematic}(b). Both the wall and vapor temperature are at the saturation temperature, \ie, $T_w=T_g=T_\text{sat}$, while the liquid is superheated and the liquid temperature at the right boundary of the domain, which is far from the interface, is fixed at $T_{\infty}$. The heat flux from the liquid to the interface results in vaporization. Similarly, the generated vapor pushes the liquid to the right. The analytical solution for the interface position has the identical expression as that for the Stefan problem, \ie, Eq.~\eqref{eq:stef_interface}, though the parameter $\beta$ is computed from a different transcendental equation
\begin{align}
	\exp{(\beta^2)} \erf{(\beta)}\left[\beta - \frac{(T_{\infty} - T_\text{sat}) C_{p,g} k_l \sqrt{\alpha_g} \exp{\left(-\beta^2 \frac{\rho_g^2 \alpha_g}{\rho_l^2 \alpha_l} \right)}}{h_{lg} k_g \sqrt{\pi \alpha_l} \erfc{ \left( \beta \frac{\rho_g \sqrt{\alpha_g}}{\rho_l \sqrt{\alpha_l}}\right)}} \right] = 0\, .
	\label{eq:suck_beta}
\end{align}
The exact solution of the liquid temperature is given as
\begin{align}
	T_l(x,t) = T_{\infty} - \left( \frac{ T_{\infty}-T_\text{sat} }{ \erfc{ \beta \frac{\rho_g \sqrt{\alpha_g}}{\rho_l \sqrt{\alpha_l}}}} \right) \erfc {\left(\frac{x}{2 \sqrt{\alpha_l t}} + \beta \frac{ \rho_g-\rho_l}{\rho_l} \sqrt{\frac{\alpha_g}{\alpha_l}} \right)}.\,
	\label{eq:suck_T}
\end{align}

%%%%%%%%%%%%%% Figure 4 %%%%%%%%%%%%%%%%%%%
\begin{figure}[tbp]
	\begin{center}
		\includegraphics [width=1\columnwidth]{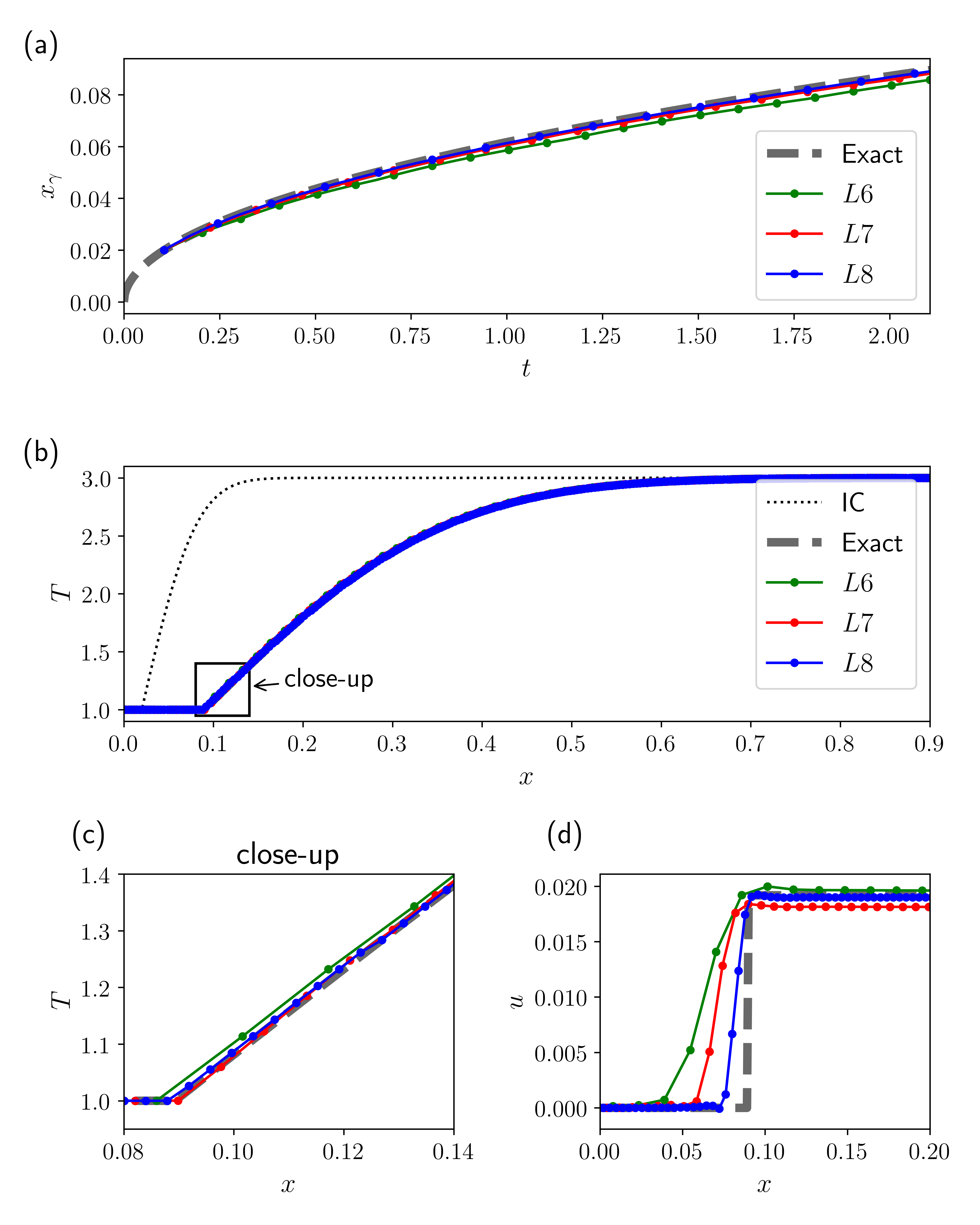}
	\end{center}
	\caption{Numerical results for the 1D Stefan problem for three levels of grid refinement: $L6$, $L7$, and $L8$, compared with the exact solution. (a) Temporal evolution of the interface position $x_\gamma$. (b)-(c) The overall temperature $T$ and (d) the $u-$velocity distributions at the final time $t=2.1\,$s. Note that (c) is a close-up of the temperature distribution near the interface - see annotation in (b).  }
	\label{fig:sucking_results} 
\end{figure}

%%%%%%%%%%%%%% Table 3 %%%%%%%%%%%%%%%%%%%
 \begin{table*}[tbp]
 \centering
 \begin{tabular}{c c c c} 
     \hline
 Maximum grid level    & Final interface location  & Relative error (\%) & $\mathcal{O}$\\
     \hline
 $L6$    & 0.086 & 4.2 & -\\
 $L7$  & 0.088 & 1.5 & 1.46\\ % convergence O = ln(E_h/E_h/2)/ln(2)
 $L8$ & 0.089 & 0.6 & 1.42\\
     \hline
 \end{tabular}
 \caption{The relative error of the interface position at the final time for the 1D sucking problem and the order of convergence  $\mathcal{O}$.}
 \label{tab:cases}
 \end{table*}

In this test case, we have set $L_0 = 1\,$m, $H_0=0.05\,$m,  and $T_{\infty}/T_{sat}=3$. The fluid properties (Fluid A) are provided in Table~\ref{tab:properties}. {The exact solution at $t=0.1$ s was used as the initial condition, and then the simulation was run to $t=2.1\,$s.} The different meshes  $L6$, $L7$, and $L8$ were used and the results are presented in Fig.~\ref{fig:sucking_results}. It is clearly shown that the numerical results converge toward the exact solution when the mesh is refined. The trajectory of the interface and final temperature profile for the mesh $L8$ match the exact solution very well. Different from the Stefan problem, the liquid temperature near the interface rises rapidly and non-linearly in $x$ in the sucking problem. 
The interface positions at the final time ($t=2.1$ s) for different meshes are provided in Table~\ref{tab:cases}, from which it can be seen that the relative error decreases with the cell size. The order of convergence is approximated using the error of the final position ($\epsilon$)
\begin{align}
\mathcal{O} = \frac{ln(\epsilon_{\Delta x}/\epsilon_{\Delta x/2})}{ln(2)}
\end{align}

Similar to the 1D Stefan problem, the distribution of $\hat{s}$ to the pure gas cells results in a non-zero vapor velocity in a narrow region on the left of the interface and as the mesh is refined, the thickness of the non-zero velocity region reduces (Fig.~\ref{fig:sucking_results}(d)). More importantly, it is clearly shown that the liquid velocity on the right of the interface converges to the exact solution. 

%===============================================================================
%   3D Bubble in a superheated liquid
%===============================================================================
\subsection{Spherical bubble growth in superheated liquid}
\label{section:bubble}
In this test, we simulate the growth of a spherical vapor bubble surrounded by the superheated liquid \cite{malan_geometric_2021,bures_direct_2021, sato_sharp-interface_2013, zhao_boiling_2022,gao_effect_2022}. The vapor is at saturation temperature $T_\text{sat}$ and the liquid temperature in the far-field is fixed at $T_{\infty}>T_\text{sat}$. The heat flux from the liquid side drives vaporization and bubble growth. The gravity effect is ignored, so the bubble remains spherical as it grows. This problem is the spherically symmetric analog of the 1D sucking problem. The analytical solution for the temporal evolution of the bubble radius ($R$) is
\begin{align}
	R(t) = 2 \beta \sqrt{\alpha_l t}, \, 
	\label{eq:bubble_radius}
\end{align}
where $\beta$ is obtained by solving the following equation, 
\begin{align}
    2 \beta^2 \int_{0}^{1} exp\left( -\beta^2 \left( (1-\zeta)^{-2} - 2(1-\frac{\rho_g}{\rho_l}) \zeta -1 \right) \right) \,d\zeta \notag\\
    = \frac{\rho_l C_{p,l}(T_{\infty}-T_\text{sat})}{\rho_g(h_{lg} + (C_{p,l} - C_{p,g})(T_{\infty}-T_\text{sat}))}\, . 
	\label{eq:bubble_beta}
\end{align}
The liquid temperature is expressed as a function of $r$ and $t$ as
\begin{align}
	T_l(r,t) = T_{\infty} - 2 \beta^2 \left(  \frac{\rho_g(h_{lg} + (C_{p,l} - C_{p,g})(T_{\infty}-T_\text{sat}))}{\rho_l C_{p,l}}  \right) \notag \\
	\int_{1-R(t)/r}^{1} exp\left( -\beta^2 \left( (1-\zeta)^{-2} - 2(1-\frac{\rho_g}{\rho_l}) \zeta -1     \right) \right) \,d\zeta\, .
	\label{eq:bubble_T}
\end{align}

\begin{figure}[tbp]
	\begin{center}
		\includegraphics [width=0.7\columnwidth]{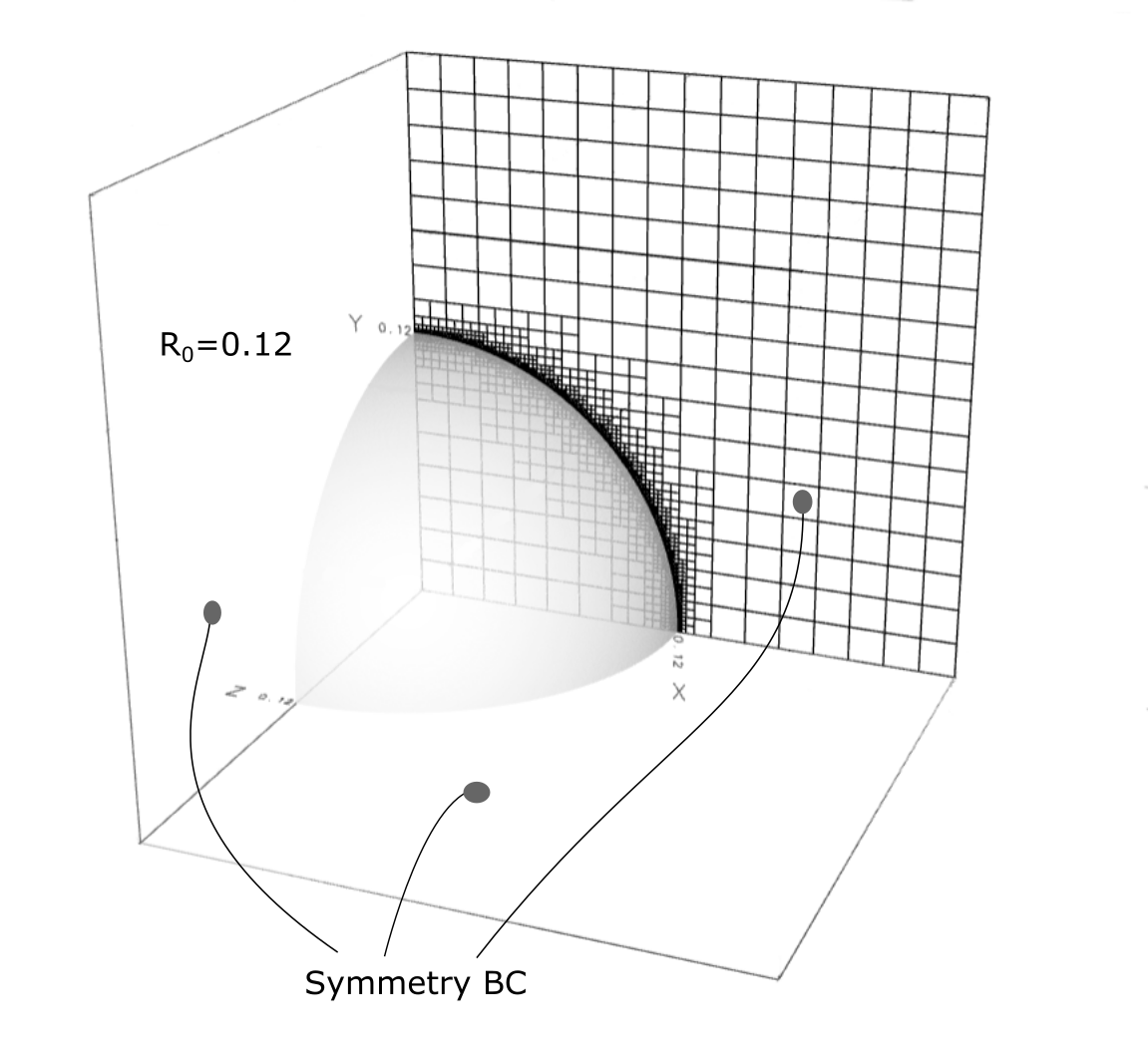}
	\end{center}
	\caption{The 3D computational domain for a spherical bubble in a superheated liquid.}
	\label{fig:3D_bubble} 
\end{figure}

Figure~\ref{fig:3D_bubble} shows the initial bubble ($R_0=1.2$ m) in the cubic computational domain with edge length $L_0=5R_0$. The octree mesh is plotted on the $x$-$y$ plane, and it can be seen that the mesh is refined to the maximum level 
near the interface to guarantee the interface and temperature gradient will be well captured. The fluid properties (Fluid A) are provided in Table~\ref{tab:properties} \cite{malan_geometric_2021}. The exact solution for $t=0.5\,$s is used as the initial condition, and the simulation is run to $t=0.21\,$s. 

%%%%%%%%%%%%%% Figure 5 %%%%%%%%%%%%%%%%%%%

\begin{figure}[tbp]
	\begin{center}
		\includegraphics [width=1.\columnwidth]{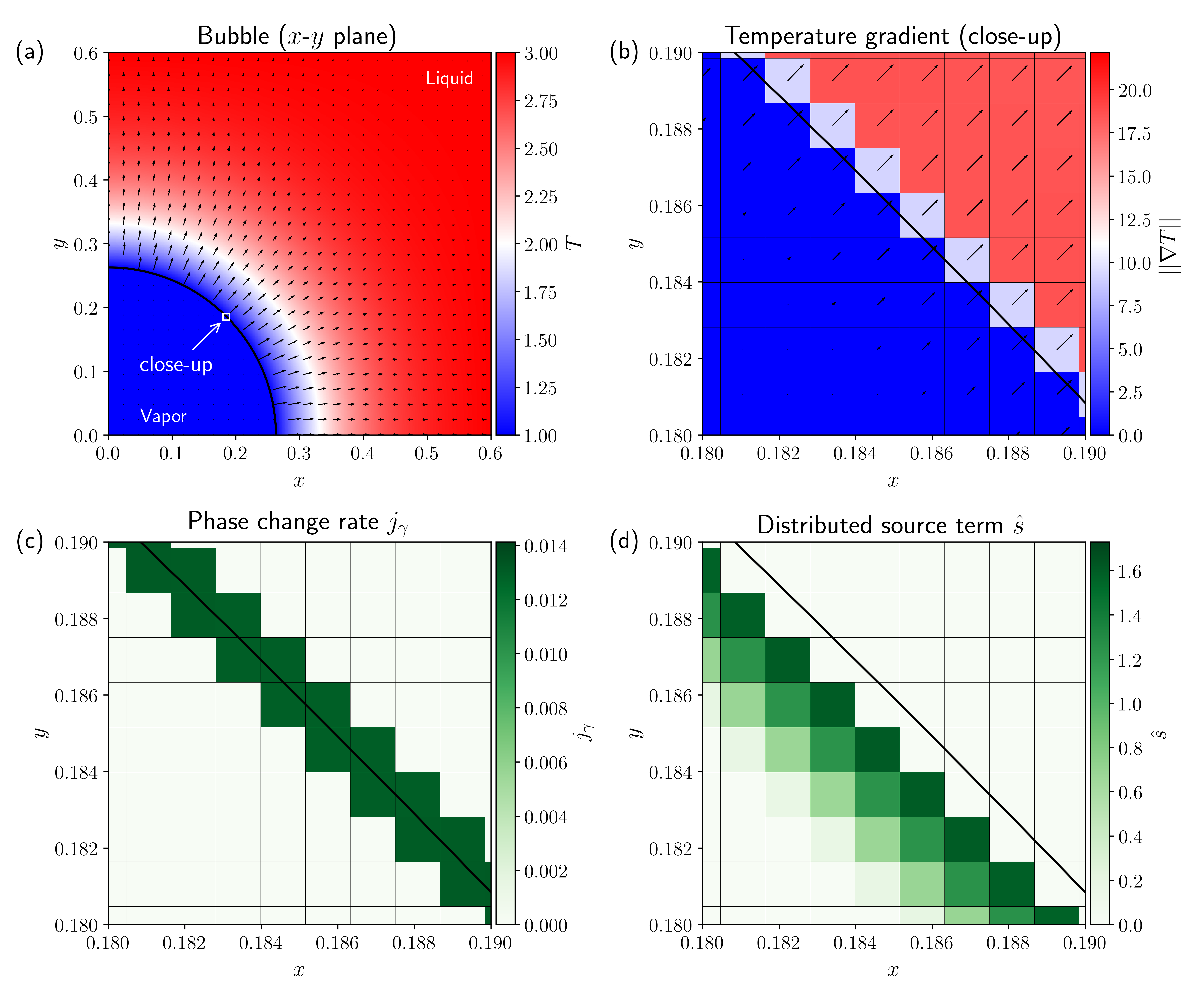}
	\end{center}
	\caption{Bubble $x$-$y$ plane cross-section at $t=0.21$ showing (a) the temperature distribution and the velocity quiver plot. Three close-up depictions of cells near the interface showing (b) the magnitude in the temperature gradient, (c) the rate of phase change at the interface ($j_{\gamma}$), and (d) the distributed phase-change source term ($\hat{s}$).}
	\label{fig:bubble_cross_section} 
\end{figure}

The temperature distribution on the central $x$-$y$ cross-section is shown in Fig.~\ref{fig:bubble_cross_section}(a). It can be seen that the bubble surface is spherical and smooth, as it is supposed to be. The temperature is fixed at $T_{sat}$ inside the bubble and increases radially outward in the liquid from the interface location. A close-up of the interface is shown in Fig.~\ref{fig:bubble_cross_section}(b), where the velocity field (vectors) and the magnitude of the temperature gradient (color) are plotted. The velocity jump across the interface can be clearly seen. More importantly, the velocity field obtained by the present method does not show artificial fluctuations of velocity magnitude or direction inside the bubble and at the interface, as observed in the previous studies (see Figs.~9-10 in Ref.~\cite{gao_effect_2022} and Fig.~8 in Ref.~\cite{malan_geometric_2021}). The rate of vaporization ($j_\gamma$), computed based on the temperature gradient, is plotted in Fig.~\ref{fig:bubble_cross_section}(c), which is non-zero only in the interfacial cells. The volumetric source $s_\gamma$ is computed from $j_{\gamma}$ using Eq.~\eqref{eq:sm} and is then distributed to the neighboring pure gas cells, and the distributed source $\hat{s}$ is plotted in Fig.~\ref{fig:bubble_cross_section}(d). It can be seen that $\hat{s}$ is smoothly distributed in the nearby pure gas cells and the gas cells close to the interface receive a bigger share. 

%%%%%%%%%%%%%% Figure 6 %%%%%%%%%%%%%%%%%%%

\begin{figure}[tbp]
	\begin{center}
		\includegraphics [width=1\columnwidth]{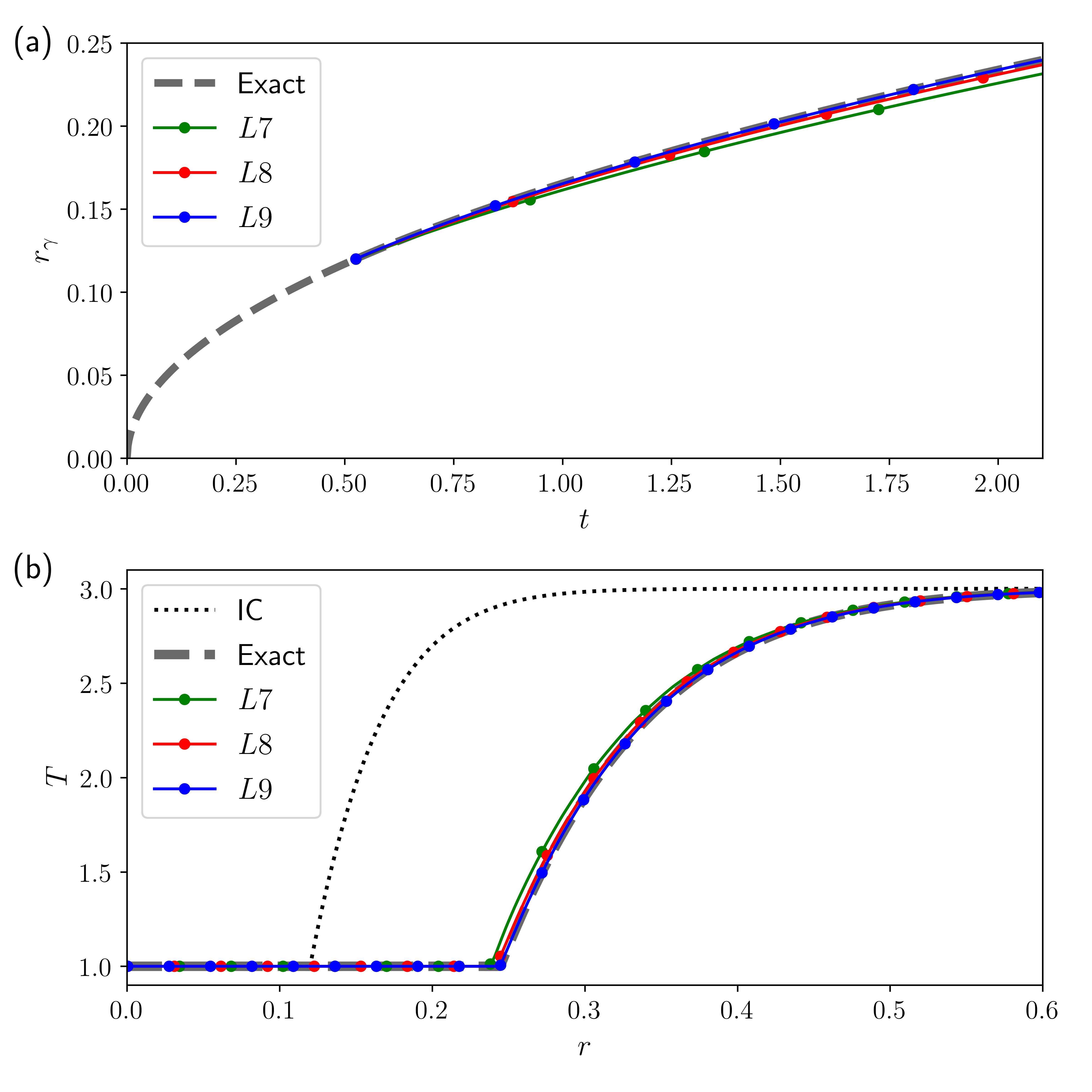}
	\end{center}
	\caption{Numerical results for the spherical bubble growth problem for three levels of grid refinement: $L7$, $L8$, and $L9$, compared with the exact solution. (a) Temporal evolution of the radial interface position $r_\gamma$. (b) Temperature distribution at the final time $t=0.21\,$s}
	\label{fig:bubble_growth_rate} 
\end{figure}

The temporal evolution of the bubble radius and the radial distribution of temperature at the final time are presented in Fig.~\ref{fig:bubble_growth_rate}. It is observed that both the numerical results converge toward the analytical solution when the mesh is refined from $L7$ to $L9$. The results for $L9$ agree with the exact solution very well. The values of the bubble radius and the relative errors at the final time for various levels of refinement are provided in Table~\ref{tab:bubble_results}. It is clear that the error is reduced when the mesh is refined. 

%%%%%%%%%%%%%% Table 4 %%%%%%%%%%%%%%%%%%%
 \begin{table*}[tbp]
 \centering
 \begin{tabular}{c c c c} 
     \hline
 Maximum grid level  & Final radius  & Relative error (\%) & $\mathcal{O}$\\
     \hline
 $L7$    & 0.231 & 3.7 & -\\
 $L8$  & 0.236 & 1.44 & 1.36\\
 $L9$ & 0.239 & 0.26 & 2.47\\
     \hline
 \end{tabular}
 \caption{The relative error of the interface position at the final time for the spherical bubble problem and the order of convergence  $\mathcal{O}$.}
 \label{tab:bubble_results}
 \end{table*}
 
%%%%%%%%%%%%%% Figure 7 %%%%%%%%%%%%%%%%%%%

%===============================================================================
%   Film boiling
%===============================================================================
\subsection{Film boiling}
\label{section:film_boil}
The last validation test case to be presented is the 2D film boiling problem \citep{berenson_film-boiling_1961, klimenko_film_1981, sun_modeling_2014, esmaeeli_computations_2004, tomar_numerical_2005, hardt_evaporation_2008}. The buoyancy effect is included which triggers the Rayleigh-Taylor instability (RTI) at the interface. Compared to the previous tests, which involve only 1D flows, this test exhibits 2D flows and more complex interface deformation. 

The most unstable wavelength for inviscid RTI with surface tension is given as 
\begin{align}
\lambda_d = 2 \pi \sqrt{\frac{3 \sigma}{(\rho_l-\rho_g)g}}\,,
\label{eq:film_wavelength}\,
\end{align}
based on this we have set the computational domain $x=[-\lambda_d/2,\lambda_d/2]$ and $y=[0,\lambda_d]$, so that 
the width of the domain covers one most-unstable wavelength. The initial interface is perturbed with a single mode and the interfacial position is expressed as 
\begin{align}
y=\frac{\lambda_d}{128}\left[ 4+cos\left( \frac{2 \pi x}{\lambda_d}\right)\right].
\label{eq:film_init}
\end{align}

%%%%%%%%%%%%%% Figure 8 %%%%%%%%%%%%%%%%%%%
 \begin{figure}[tbp]
	\begin{center}
		\includegraphics [width=0.7\columnwidth]{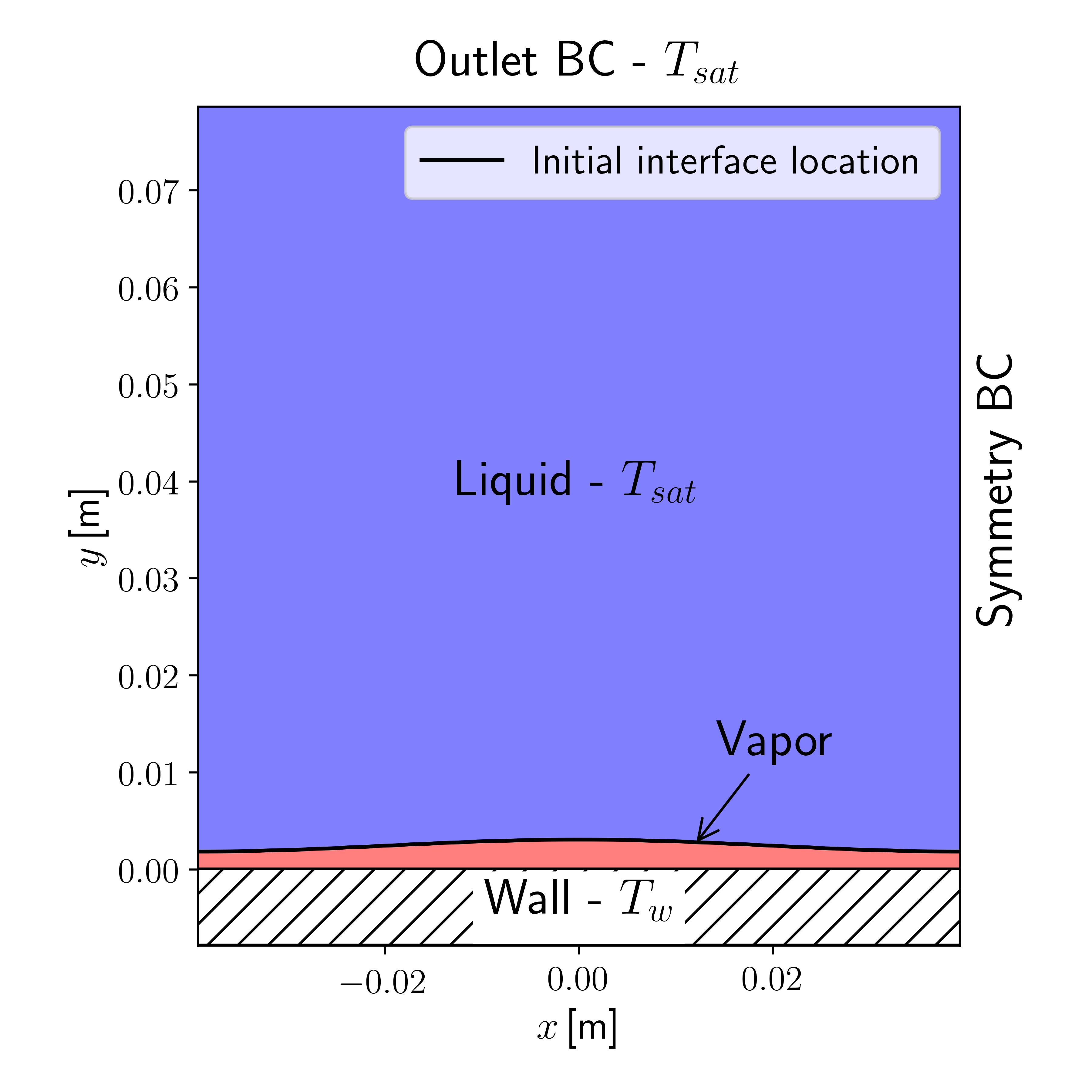}
	\end{center}
	\caption{The schematic of the computational domain for the film boiling problem.}
	\label{fig:film_domain} 
\end{figure}

The fluid properties are provided in Table~\ref{tab:properties} (Fluid B) and the gravitational acceleration is $g=9.81\,$m/s. The bottom wall is a no-slip boundary at a constant elevated temperature $T_{wall}=T_\text{sat}+5\,K$. The left and right boundaries are symmetry boundary conditions and the top boundary is an outlet boundary condition. Initially, the fluids are stationary and the temperature is varied linearly in the vapor between the wall and the interface. 

%%%%%%%%%%%%%% Figure 9 %%%%%%%%%%%%%%%%%%%
\begin{figure}[tbp]
	\begin{center}
		\includegraphics [width=1.0\columnwidth]{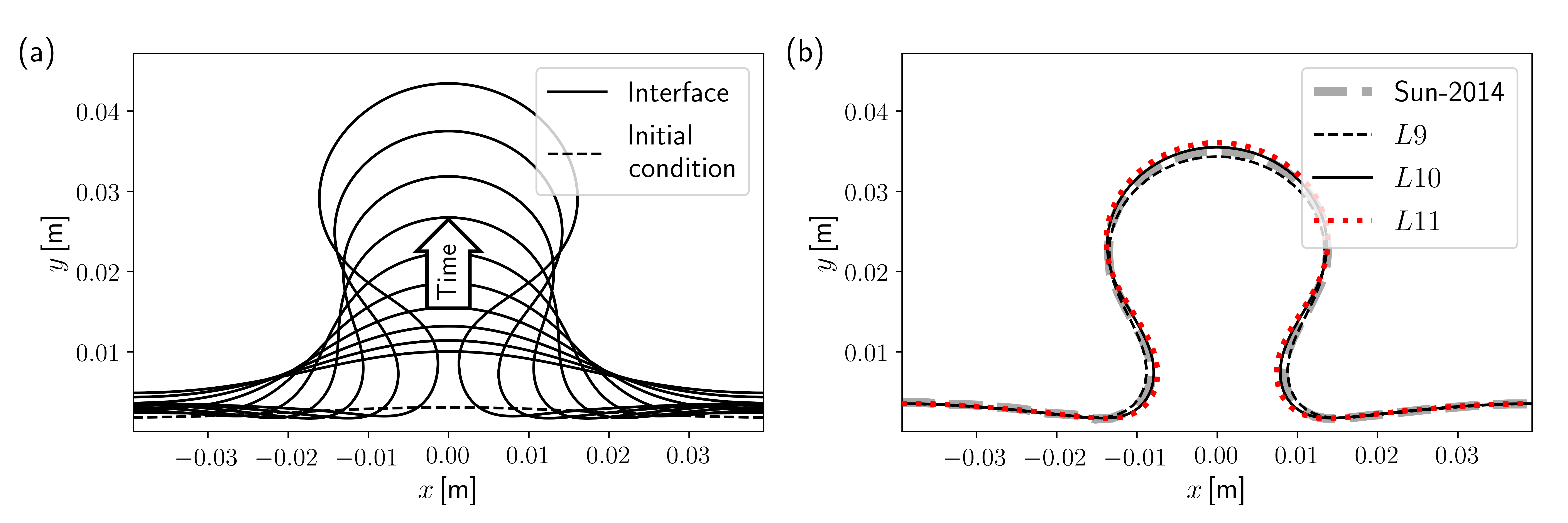}
	\end{center}
	\caption{(a) The temporal evolution of the interface from $t=0.237\,s$ to $t=0.453\,s$ at increments of $0.024\,s$, and (b) the interface shape of the film boiling problem at three levels of refinement ($L9$, $L10$, and $L11$) compared to the results from Sun \etal \citep{sun_modeling_2014} at $t\approx 0.24s$.}
	\label{fig:film_converge} 
\end{figure}

Figure~\ref{fig:film_converge}(a) shows the temporal development of the interface from $t=0.237\,$ to $0.453\,$s at a constant increment of $0.024\,$s. The development of the RTI bubble can be clearly seen. In Fig.~\ref{fig:film_converge}(b), the present results at $t\approx 0.24s$ for different mesh refinement levels are compared to the numerical results of Sun \ etl \citep{sun_modeling_2014}. Again,  the present numerical results converge as the mesh is refined, and the results for L11 are in good agreement with the previous numerical results using a uniform mesh. The solution appears to have converged at $L10$ as the difference between results for $L10$ and $L11$ is almost invisible (Fig.~\ref{fig:film_converge}).

\section{Deformation and breakup of a vaporizing drop}
The results for the above validation cases and the good agreement with exact solutions and previous numerical studies have affirmed that the present methods and the implementation in the \emph{Basilisk} code are capable of accurately resolving interfacial liquid-gas two-phase flows
with vaporization. In this section, the code will be applied to simulate the aerodynamic breakup of a vaporizing drop in a uniform hot gas stream. 

It is considered here that  a spherical drop at saturated temperature ($T=T_\text{sat}$) is stationary at $t=0$ 
and is suddenly exposed to an unbounded uniform hot vapor stream ($T=T_\infty$) at $t=0^+$. 
When vaporization is absent ($T_\infty= T_\text{sat}$), the drop dynamics and topology evolution 
are fully determined by the densities and viscosities of the drop liquid and the gas, 
$\rho_l,\ \mu_l,\ \rho_g,\ \mu_g$, the surface tension $\sigma$, the initial drop diameter $D_0$, 
and the uniform gas stream velocity $U_\infty$. The subscript $0$ is used to represent the initial state. 
These parameters lead to four independent dimensionless parameters: 
the Weber number, $\text{We}=\rho_g U_\infty^2 D_0/\sigma$, the Reynolds number, $\text{Re}=\rho_g U_\infty D_0/\mu_g$, 
the Ohnesorge number, $\text{Oh}=\mu_l/\sqrt{\rho_l D_0 \sigma}$, 
and the gas-to-liquid density ratio, $\eta=\rho_g/\rho_l$ \citep{pilch_use_1987, hsiang_near-limit_1992, joseph_breakup_1999, guildenbecher_secondary_2009}.  Alternative dimensionless parameters can be defined based on the above four parameters \citep{guildenbecher_secondary_2009}. For drops with low Oh, such as millimeter drops of low-viscosity liquids like water, the viscous stress is small compared to surface tension and thus the latter is the dominant stabilization mechanism against drop deformation and breakup. In such cases, $\text{We}$ is the most important parameter and is typically used to characterize the breakup modes. For low-Oh drops, the critical Weber is number $\text{We}_{cr}=11\pm2$ \cite{guildenbecher_secondary_2009, theofanous_physics_2012}, and the drop will break only when $\text{We}> \text{We}_{cr}$. Recent detailed numerical simulations of drop aerobreakup indicated that 2D axisymmetric simulations yield good approximation only for non-breaking drops at low We and Re \cite{mahmood_effects_2021}. For drops with high We and Re, which will break in a bag or multi-bag mode, fully 3D simulations are required to accurately capture the drop dynamics. Here we consider two cases of different We. 
In the first case, we considered a water drop at $\text{We}=1.5$ and $20<\text{Re}<200$. Since $\text{We}$ is significantly lower than $\text{We}_{cr}$, the drop will not break and the deformation is mild, so 2D axisymmetric simulations were performed. In the second case, we have considered an acetone drop at $\text{We}=62.7$. The drop will break and thus a fully 3D simulation was conducted. 

When $T_\infty> T_\text{sat}$, vaporization will occur and the rate is controlled by the Stefan number (also referred to as the Spalding  and Jakob numbers), $\text{St}= C_{p,g}(T_{\infty}-T_\text{sat})/h_{lg}$. In both cases considered here, $\text{St}$ is about 0.1. When the drop deformation is not negligible, the increase in the interfacial area will result in an increase in the drop vaporization rate. In such cases, the heat and mass transfer between the drop and the surrounding vapor will deviate from the models for spherical drops \cite{renksizbulut_experimental_1983}, which is strictly valid for zero We.
 
Sub-grid drop vaporization models for drops are important to the accurate simulation of sprays in practical applications. For sprays consisting of a large number of drops, it is inviable to resolve the interface for each individual drop. Instead, a drop is modeled as a point mass (or a group of drops as a parcel), which is known as the Lagrangian point-particle (LPP) approach \cite{balachandar_scaling_2009}. Since the flow and temperature fields in the drop scale are not resolved, physical models are required to account for the momentum, heat, and mass transfer between the drop and the surrounding gas and to predict the children droplets generated after breakup \cite{wert_rationally-based_1995, dai_temporal_2001, kuo_maximum_2022}.  For drops with finite We, the drop deformation/breakup and vaporization are closely coupled.  LPP models that can accurately capture this coupling effect remain to be established. High-fidelity detailed numerical simulations, enabled by the present method, can provide crucial insights into the currently unclear physics and, thus, are important to the development of such models in the future. 

%===============================================================================
%   Moving drop evaporation
%===============================================================================
\subsection{Axisymmetric 2D simulations for a vaporizing drop at low We}
\label{section:droplet}
We first consider a water drop at a low Weber number. The water properties are provided in Table~\ref{tab:properties} where the Prandtl number of water in steam is $\text{Pr}\approx1$. The drop is initially spherical 
with a radius $R_0$. A parametric study has been performed by varying the Reynolds number, \ie, $22<\text{Re}<200$, 
and there are in total 25 cases simulated. For each case, we have modified the free stream velocity $U_\infty$ and surface tension ($\sigma$) simultaneously to keep the Weber number fixed at $\text{We}=1.5$. Due to the low We and Re, the drop deformation is mild. Therefore, 2D axisymmetric simulations are sufficient to capture the drop dynamics and vaporization. For all cases, the free-stream temperature is set to $T_{\infty}\approx600\,$K, which yields $\text{St}=0.1$. 

%%%%%%%%%%%%%% Figure 10 %%%%%%%%%%%%%%%%%%%
 \begin{figure}[tbp]
	\begin{center}
		\includegraphics [width=0.9\columnwidth]{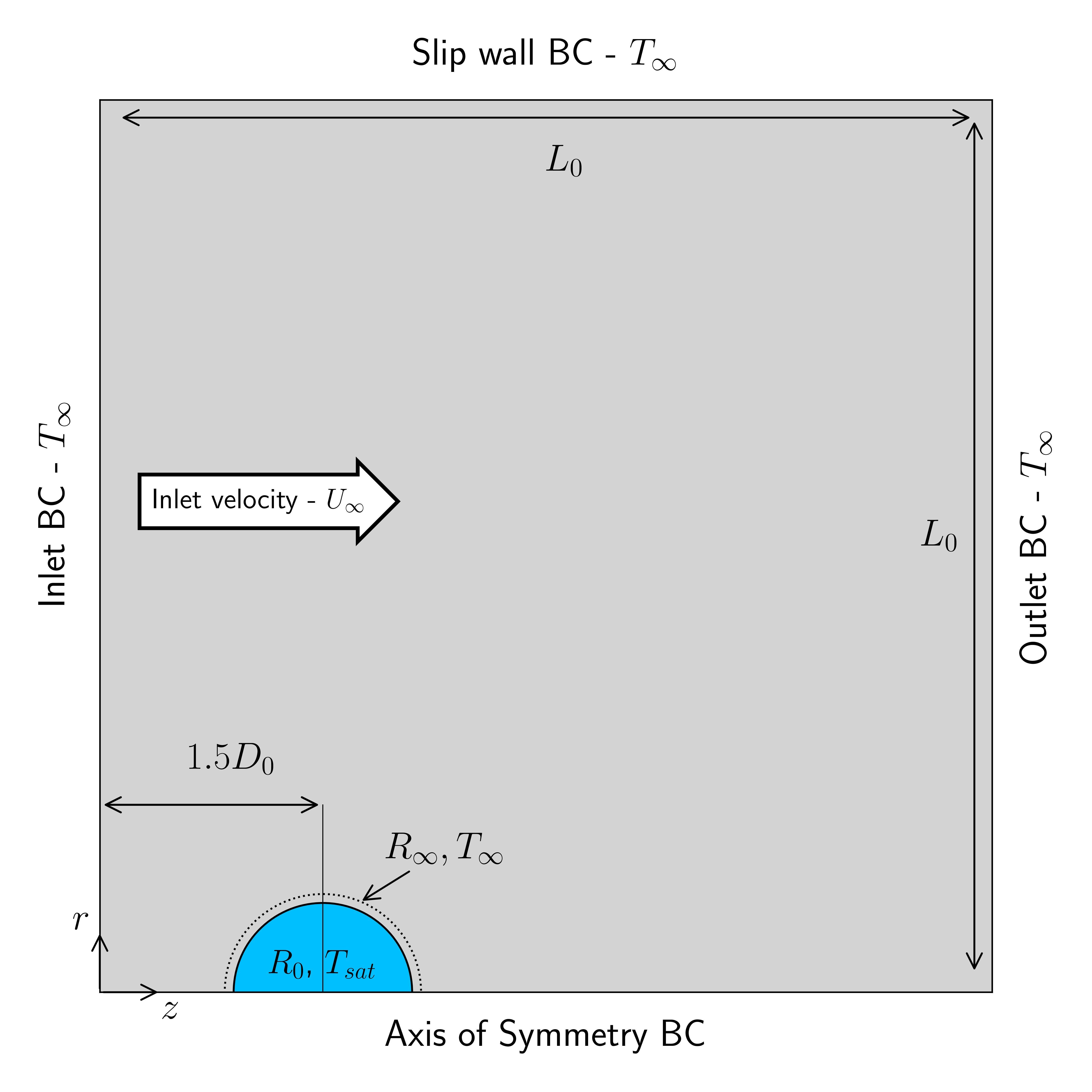}
	\end{center}
	\caption{The computational domain for 2D-axisymmetric simulation of the vaporization of a freely moving
	drop in a hot vapor stream.}
	\label{fig:moving_droplet_diagram} 
\end{figure}

The computational domain for the 2D axisymmetric simulation is shown in Fig.~\ref{fig:moving_droplet_diagram}.
The domain is a square with the edge length $L_0=8D_0$. The drop is initially located at $x=1.5D_0$. For all cases, we have run the simulations with $L12$ (equivalent to $\Delta x = D_0/512$) to $t^*=0.16$, unless stated otherwise. Note that $t^*$ is the dimensionless time defined as \cite{ranger_aerodynamic_1969}
\begin{align}
t^* = t\frac{U_{\infty}}{D_0\sqrt{\rho_l/\rho_g}}.
\label{eq:t_star}
\end{align}

Within this short time duration, the drop velocity change is very small due to the large density contrast between the liquid and vapor. As a result, the relative velocity can be approximated as unchanged, so Re and We remain to be their initial values. After a short time duration for the velocity and thermal boundary layer to develop from the initial condition, the drop vaporization rate reaches an approximate quasi-steady state, until then we will measure its value. To shorten the transition to the quasi-steady state, the initial condition for the temperature is specified as
 \begin{align}
    T(r)= 
\begin{cases}
    T_\text{sat}\,,& \text{if } r\leq R_0\\
    T_{\infty}\,,& \text{if } r\geq R_{\infty}\\
    \frac{R-R_0}{R_{\infty}}(T_{\infty}-T_\text{sat})+ T_\text{sat}\,,& \text{otherwise}
\end{cases}
 	\label{eq:init_temp_drop}
\end{align}
where we set {{$R_{\infty}=1.25R_0$}} as shown in Fig.~\ref{fig:moving_droplet_diagram}. As long as the artificial thermal boundary layer thickness is small, $R_{\infty}-R_0\ll R_0$, the specific value of $R_{\infty}$ is immaterial and will not influence the reported results. For all the cases simulated, a quasi-steady state was reached before $t^*=0.06$. 

Extensive experimental studies have been conducted for the vaporization of a spherical drop at a quasi-steady state,
from which the empirical relations for the Nusselt (Nu) number have been extracted \citep{renksizbulut_experimental_1983, sazhin_advanced_2006, haywood_detailed_1989, chiang_numerical_1992, yuen_heat-transfer_1978}. 
A commonly used empirical model is the one developed by Renksizbulut and Yuen \citep{renksizbulut_experimental_1983}, which can be expressed as
\begin{align}
	Z_f \equiv  [Nu(1 + \text{St})^{0.7} - 2]\text{Pr}_f^{-{1}/{3}} = 0.57 \text{Re}_f^{{1}/{2}} \,
	\label{eq:empirical}
\end{align}
where $\text{Re}_f=2R\rho_{g,\infty} |\ub_{\infty}-\ub_d|/\mu_{f}$ and $\text{Pr}_f=C_{p,f} \mu_{g,f}/ k_f$ are the Reynolds and Prandtl numbers based the film properties. The subscript $f$ indicates that the parameters for the gas film surrounding the drop 
($T_f=(T_{\infty}+T_\text{sat})/2$). The variation in gas properties due to temperature variation in the film is ignored, so $\text{Re}_f=\text{Re}_{\infty}$.  The drop vaporization experiments were conducted for a range of Reynolds, Stefan, and Prandtl numbers:
$25<\text{Re}<2000$, $0.07<\text{St}<2.79$, and $0.7<\text{Pr}<1$ \citep{renksizbulut_experimental_1983}. Equation \eqref{eq:empirical} is therefore valid within these parameter ranges. Furthermore, it is considered the drop temperature is initially at $T_\text{sat}$, and heat transfer by radiation is neglected. Based on the convective heat transfer characterized by Nu, the rate of change of drop volume can then be computed as
 \begin{align}
	\dot{V_l} = \frac{- \dot{q} A_s }{\rho_l h_{lg}}\,,
	\label{eq:volume_evap_rate}
\end{align}
 where the drop surface area is $A_s=4\pi R_0^2$, the convective heat transfer coefficient is $h=\text{Nu}k_g/(2R_0)$,
and the rate of heat transfer is $\dot{q}=h(T_{\infty}-T_\text{sat})$. 

\begin{figure}[tbp]
	\begin{center}
		\includegraphics [width=1.\columnwidth]{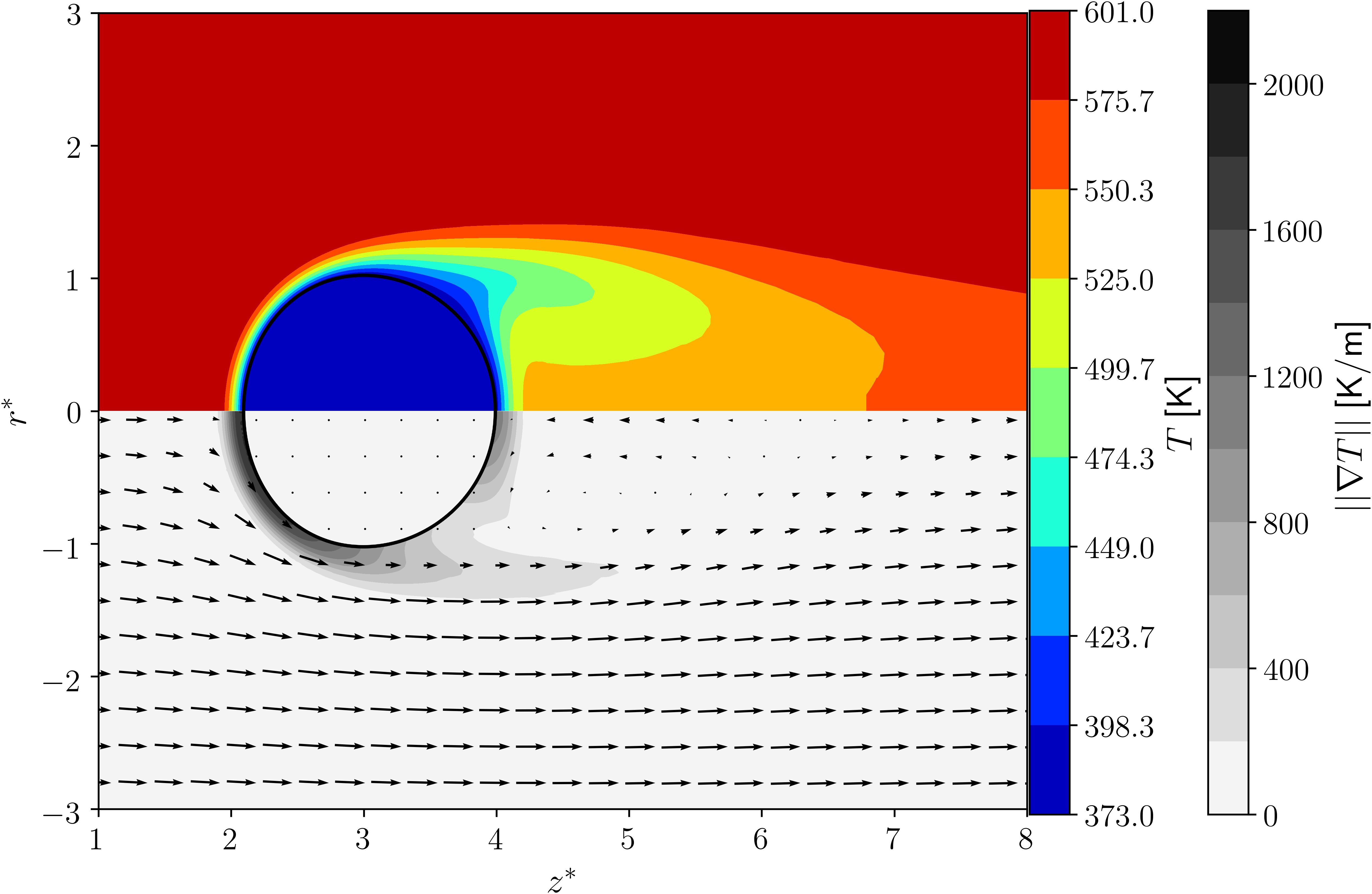}
	\end{center}
    	\caption{Temperature (top), the temperature gradient magnitude, and velocity field (bottom) for a moving water drop in high-temperature vapor steam at $t^*=0.16$ for $\text{We}=1.5$ and $\text{Re}=200$. The black contour line represents the drop surface. For better visualization, the velocity vectors are only plotted in the mesh refinement level 6. Note that $r^*=r/R_0$ and $z^*=z/R_0$.}
	\label{fig:moving_drop_temperature} 
\end{figure}

Figure~\ref{fig:moving_drop_temperature} shows the temperature and velocity fields at the end of the simulation ($t^*=0.16$) for $\text{Re}= 200$. As expected, the temperature gradient is higher on the windward surface of the drop (Figure~\ref{fig:moving_drop_temperature}), as a result, the majority of the vaporization occurs near the front stagnation point \cite{renksizbulut_numerical_1983}. On the leeward side of the drop, the temperature gradient is much lower since the gas temperature is low in the wake, where the gas is cooled by the low-temperature drop.
 
%%%%%%%%%%%%%% Figure 11 %%%%%%%%%%%%%%%%%%%
 
\begin{figure}[tbp]
	\begin{center}
		\includegraphics [width=1.\columnwidth]{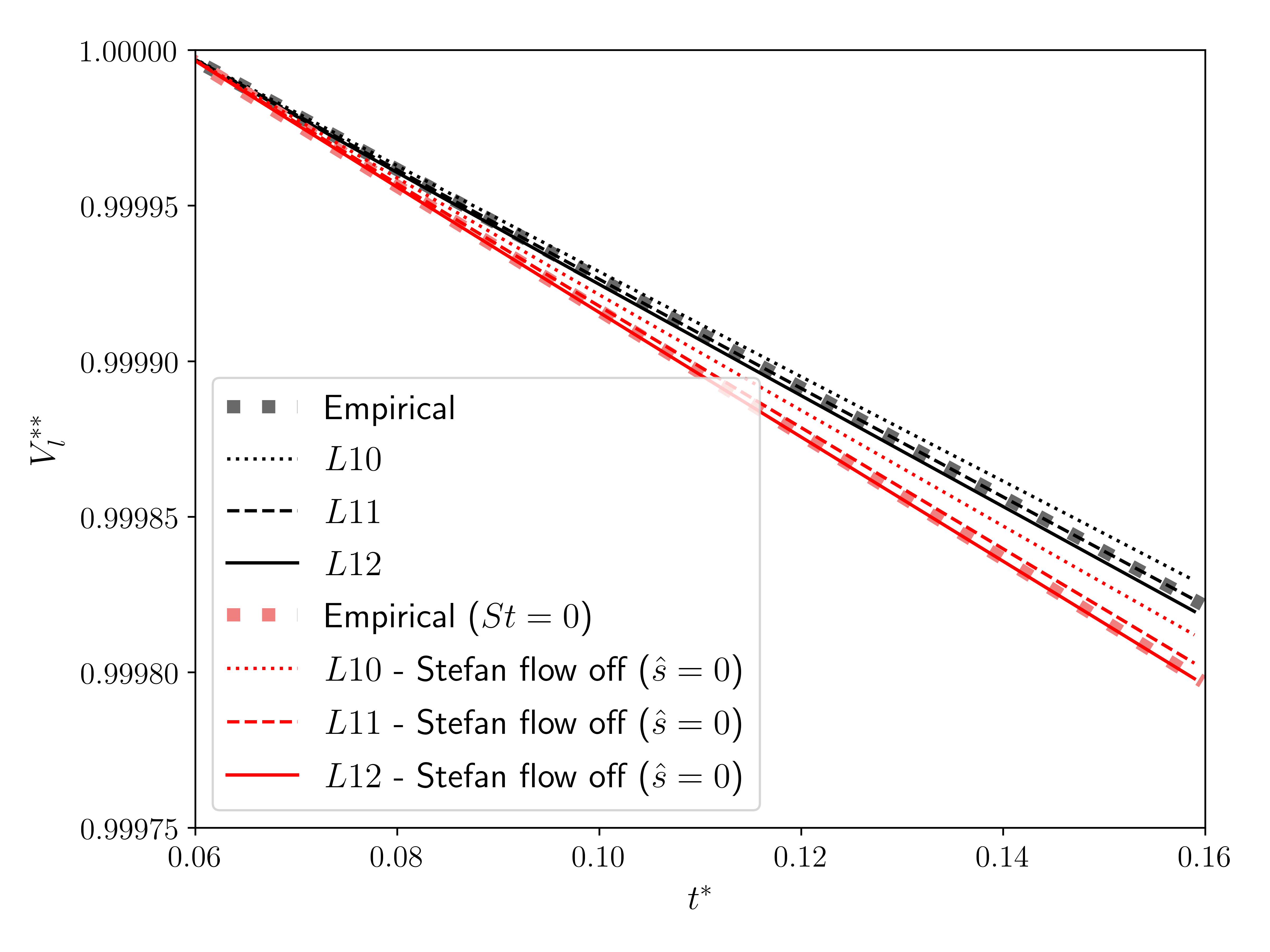}
	\end{center}
	\caption{Simulation results for the decrease of drop volume due to vaporization, with and without the Stefan flow, for \text{Re}=126. The corresponding empirical correlations are also shown for comparison. To visualize the rate of liquid volume loss, the droplet volume is non-dimensionalized using $V_l^{**}=V_l/V_l(t^*=0.06)$, where the quasi-steady state is achieved before $t^*=0.06$.}
	\label{fig:moving_drop_result} 
\end{figure}

The simulation results for the temporal evolution of the drop volume are shown in Fig.~\ref{fig:moving_drop_result}, compared with the empirical model (Eq.~\eqref{eq:volume_evap_rate}). The simulation results \emph{without} the Stefan flow are also shown to demonstrate the effect of the Stefan flow on the rate of vaporization. The Stefan flow is turned off by manually setting $\hat{s}=0$ in Eq. \eqref{eq:poisson}. Note we still consider the vaporization when we turn off the Stefan flow; \ie, $\hat{s}=0$ and $s_{\gamma}\neq0$, resulting in a reduction in the droplet volume (Fig.~\ref{fig:moving_drop_result}). The empirical model without the Stefan flow will simplify Eq.~\eqref{eq:empirical} by setting $\text{St}=0$ \citep{sazhin_advanced_2006}. For both cases, with and  without Stefan flow, the simulation results converge as the refinement level increases from $L10$ to $L12$. The simulation results for $L12$ and the empirical models agree very well. The Nusselt number measured from the simulations is compared with the predictions of the empirical model in Table~\ref{tab:nu_droplet} and an excellent agreement is observed. The Stefan flow of lower-temperature vapor (at $T_{sat}$) will reduce the temperature gradient at the interface and thus will result in a decrease in the rate of heat and mass transfer. Therefore, ignoring the Stefan flow will lead to an overestimation of the rate of drop vaporization, therefore, it is important to accurately capture the Stefan flow in a simulation. 

%%%%%%%%%%%%%% Figure 12 %%%%%%%%%%%%%%%%%%%

%%%%%%%%%%%%% Table 5 %%%%%%%%%%%%%%%%%%%
 \begin{table*}[tbp]
 \centering
 \begin{tabular}{c c c } 
     \hline
 Case    & Nu & Nu (without Stefan flow) \\
     \hline
 L10       & 7.12 & 7.84                    \\
 L11       & 7.38 & 8.22                    \\
 L12       & 7.53 & 8.43                    \\
 Empirical & 7.42 & 8.44                    \\
     \hline
 \end{tabular}
 \caption{The Nusselt numbers with and without Stefan flow for various levels of grid refinement ($L10$, $L11$, $L12$) compared to the empirically predicted Nu.}
 \label{tab:nu_droplet}
 \end{table*}

\begin{figure}[tbp]
	\begin{center}
		\includegraphics [width=0.8\textwidth]{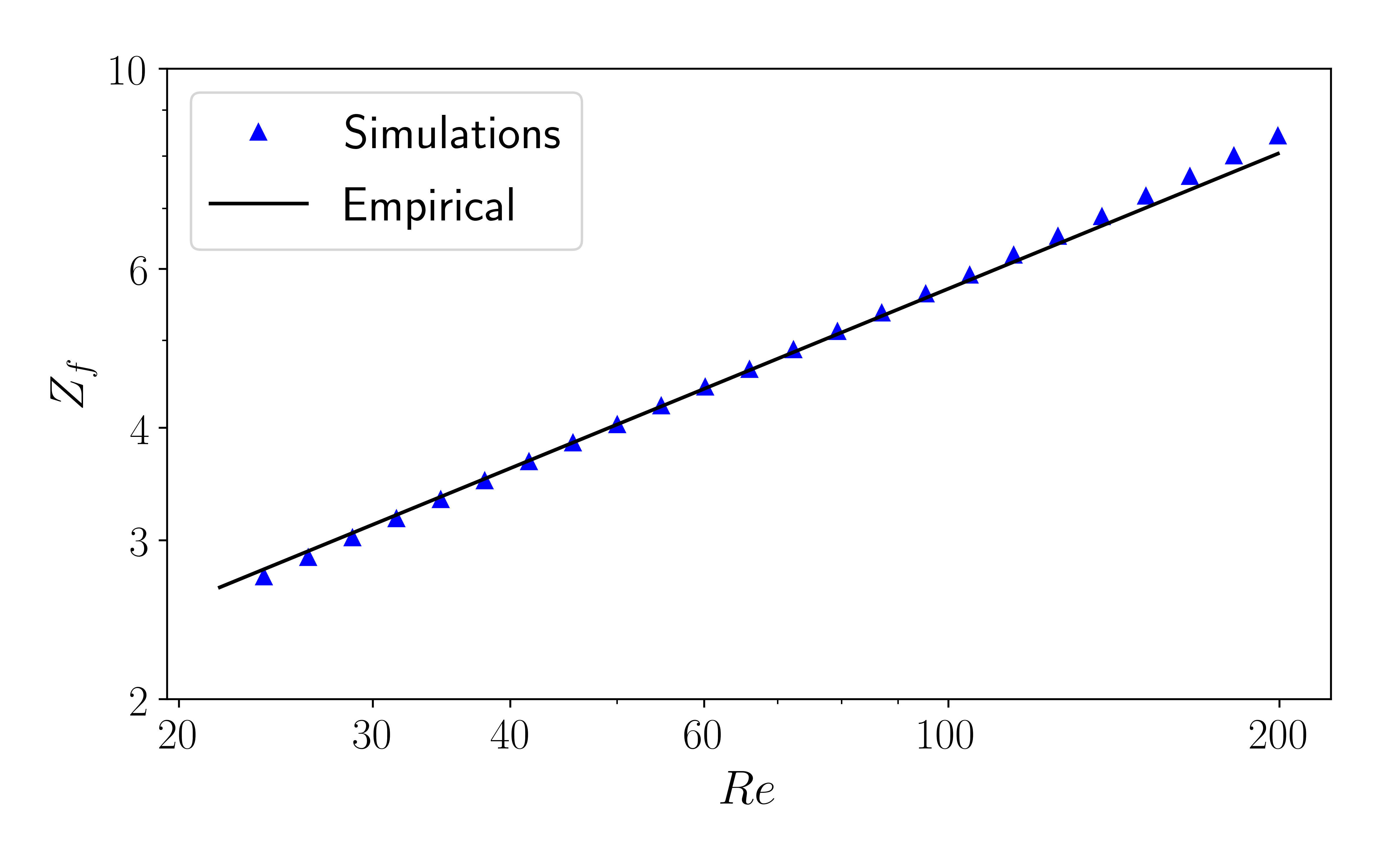}
	\end{center}
	\caption{Simulation results for different $\text{Re}$ between 20 and 200, compared with the empirical correlation ($Z_f$ is defined in Eq.~\eqref{eq:empirical}).}
	\label{fig:empir_vs_sim} 
\end{figure}

Finally, the simulation results for different Re (with mesh $L11$)  are shown in Fig.~\ref{fig:empir_vs_sim}, compared with the empirical correlation (Eq.~\eqref{eq:empirical}). The parameter $Z_f$ in Eq.~\eqref{eq:empirical} scales with $\text{Re}^{1/2}$, and the simulation results agree well with this scaling relation. The computed values of $Z_f$ match quite well with the empirical model for the whole range of Re considered. The small discrepancy may be due to the small drop deformation in the simulations since we did not constrain the drop to be perfectly spherical as in the experiment.  The good agreement observed here further validates the present methods in simulating the vaporization of a slightly deforming drop.

%%%%%%%%%%%%%% Figure 13 %%%%%%%%%%%%%%%%%%%

\subsection{Fully 3D simulation for the breakup of a vaporizing drop}
\label{3D_droplet}
\begin{figure}[tbp]
	\begin{center}
		\includegraphics [width=0.7\columnwidth]{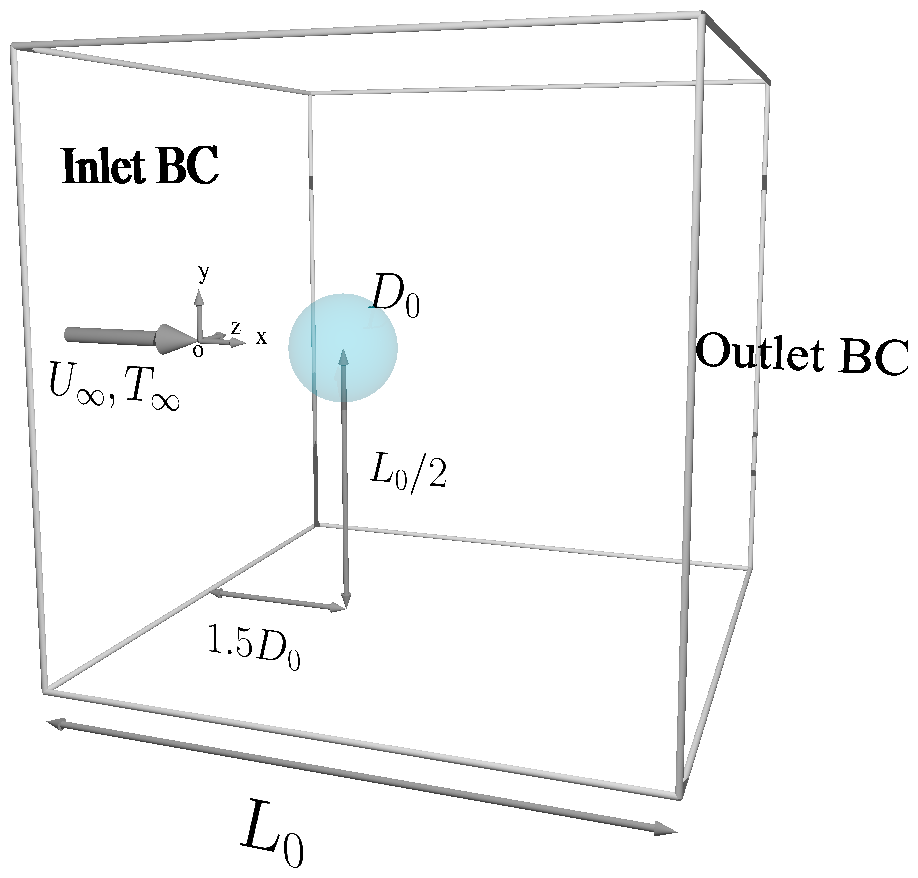}
	\end{center}
	\caption{Schematic of the 3D computational domain for 3D acetone drop exposed to a uniform high-temperature vapor stream.}
	\label{fig:3D_moving_droplet_diagram} 
\end{figure}

The value of We has been increased in the second case, as a result, the vaporizing drop deforms significantly and eventually breaks. The drop fluid is acetone and the initial diameter is 4.1 $\mu m$. The free-stream gas is acetone vapor. The liquid and vapor properties are provided in Table~\ref{tab:properties_acetone}. The free-stream gas velocity, pressure, and temperature are $U_\infty=214.4$ m/s, $p_\infty=285$ kPa, and $T_\infty=402$ K, respectively. These parameters are chosen to be similar to the post-shock conditions for a planar shock wave with Mach number 1.6, inspired by the shock tube experiment of Duke-Walker \etal \cite{duke-walker_evaporation_2021}, though the present simulation has neglected the effects of shock-drop interaction and compressibility. A 3D simulation is performed in this case and Fig.~\ref{fig:3D_moving_droplet_diagram} shows the cubic computational domain, the edge size of which is $L_0=16D_0$. The origin is at the center of the left surface of the domain. The initial position of the drop is $(x,y,z)=(1.5D_0,0,0)$. The simulation has been run to $t^*=2$.

%%%%%%%%%%%%%% Table 6 %%%%%%%%%%%%%%%%%%%
\begin{table*}[tbp]
\centering
\begin{tabular}{l l l l l} 
\hline
Property & \multicolumn{2}{c}{Acetone}  \\
&     Liquid & Vapor \\
\hline
$\rho$ $[kg/m^3]$ & $710$ & $5.11$ &  \\
$k$ $[W \, m^{-1} \, K^{-1}]$ & $0.156$  & $0.0166$\\
$C_{p,g}$  $[J \, kg^{-1} \, K^{-1}]$  & $2420$ & $1460$ \\
$\mu$ $[Pa \, s]$  & $1.85 \times 10^{-4}$ & $9.59\times 10^{-6}$\\
$h_{lg}$  $[J \, kg^{-1}]$ & $4.88 \times 10^5$ &- \\
$T_\text{sat}$  $[K]$ & $359$ &- \\
$\sigma$ $[N \, m^{-1}]$ & $0.0153$ &- \\
\hline
\end{tabular}
\caption{Acetone fluid properties.}
\label{tab:properties_acetone}
\end{table*}

%%%%%%%%%%%%%% Figure 14 %%%%%%%%%%%%%%%%%%%

 To simplify the analysis, the drop is assumed to be initially at $T_\text{sat}$ and the variation in the gas properties due to temperature is ignored. The key non-dimensional parameters are $\text{We}=62.7$, $\text{Re}=468$, $\text{Oh}=0.0277$, $\eta=0.0072$, $\text{Pr}=0.844$, and $\text{St}=0.127$. The time step is determined based on $\text{CFL}=0.5$. The maximum level of refinement is $L=12$, resulting in a minimum cell dimension $\Delta x=D_0/256$.  The total number of octree cells is approximately 11 million (equivalent to 69 billion uniform Cartesian cells). The simulation has been run on the Anvil machine for 3 days using 256 cores (AMD Epyc Milan processor). 

\begin{figure}[htp]
	\begin{center}
		\includegraphics [width=0.8\columnwidth]{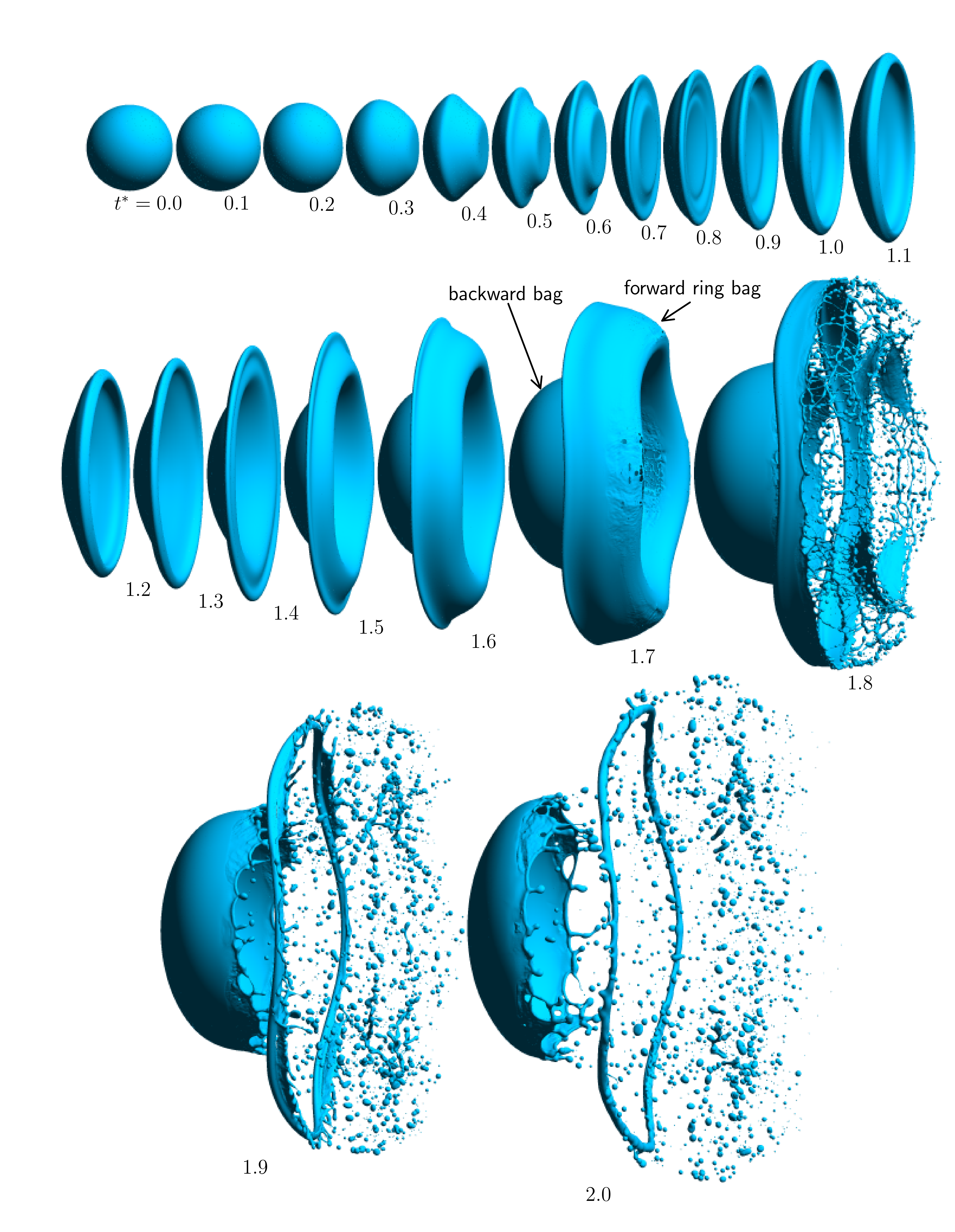}
	\end{center}
	\caption{Morphological evolution of the vaporizing drop from $t^*=0$ to $2$ at time increments of $\Delta t^*=0.1$.}
	\label{fig:acetone_drop} 
\end{figure}

\begin{figure}[htp]
	\begin{center}
	    \includegraphics [width=0.99\columnwidth]{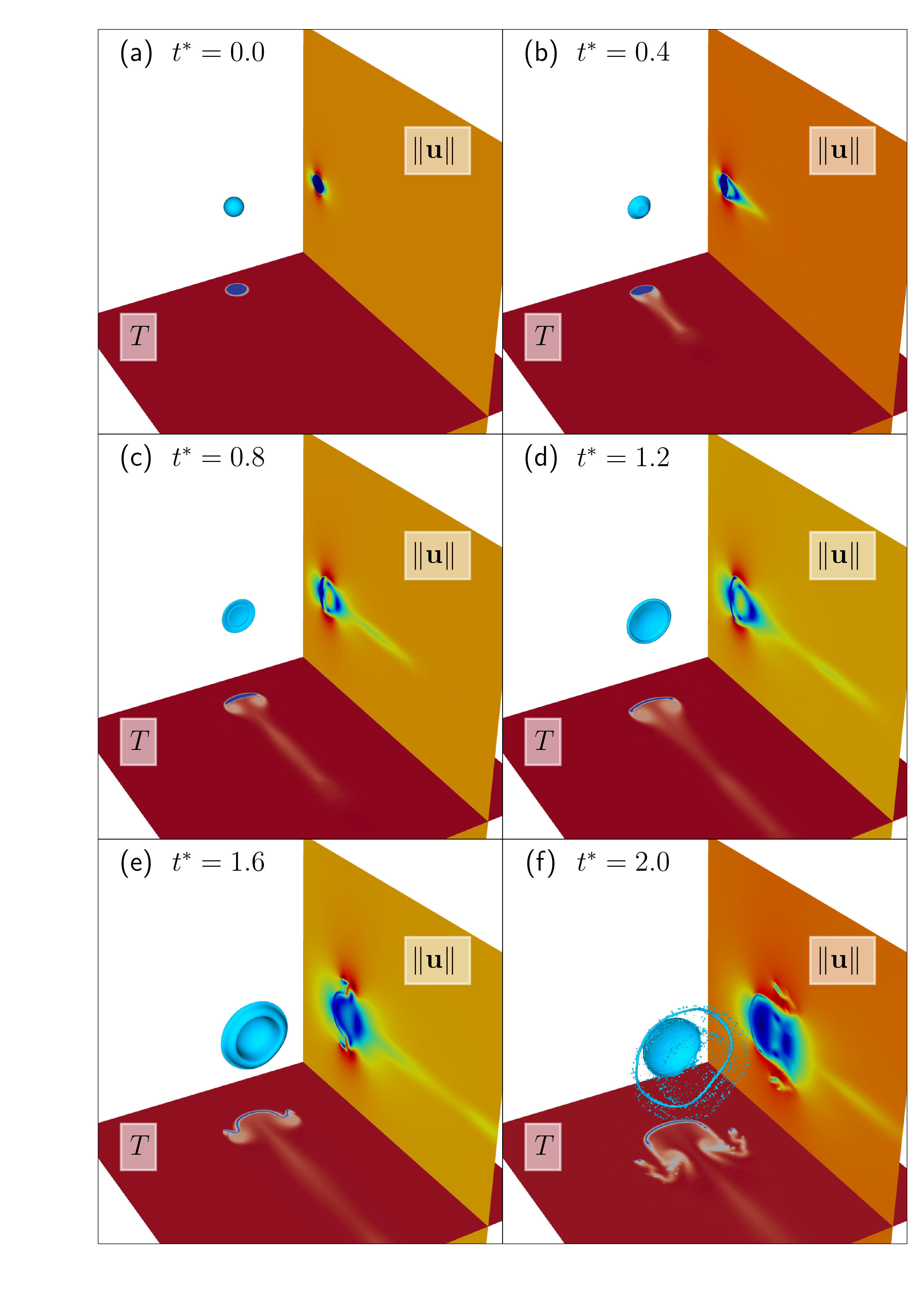}
	\end{center}
	\caption{Snapshots of the drop surface, the temperature ($T$) on the central $x$-$y$ plane, and the velocity magnitude ($\| \mathbf{u} \|$) on the central $y$-$z$ plane from $t^*=0$ to $2$ at time increments of $\Delta t^*=0.4$. }
	\label{fig:acetone_drop_u_T} 
\end{figure}

The deformation and breakup of the vaporizing acetone drop are shown in Fig.~\ref{fig:acetone_drop}. The time snapshots are from $t^*=0$ to 2.0 with an increment of $0.1$. To better illustrate the interaction between the drop and the surrounding vapor, Fig.~\ref{fig:acetone_drop_u_T} shows the drop surface along with the contours of the temperature $(T)$ and velocity magnitude $(\| \mathbf{u} \|)$ from the central $x$-$z$ and $x$-$y$  planes.

In a very short time after the initially stationary drop is exposed to the gas stream, the viscous and capillary effects are negligible and the flow field is dominated by the inviscid mechanisms. The two stagnation points can be seen on both the windward and leeward poles of the drop. As a result, the drop is compressed in the streamwise direction, and the shape of the drop is approximately symmetric with respect to the central $y$-$z$ plane normal to the streamwise direction, see $t^*=0$ to 0.2. As time evolves, the boundary layer separates and the wake starts to form. The leeward side of the drop becomes flat ($t^*=0.3$ to 0.5). The streamwise compression of the drop results in lateral expansion and an ``edge" of high curvature is formed at the periphery. The high gas velocity at the periphery results in a strong shear, and the shear Kelvin-Helmholtz (KH) like instability drives the edge roll over the back of the drop, turning the drop to a bowl with the opening facing downstream (sometimes also referred to as a backward bag) ($t^*=0.6$ to 1.3). As the drop continues to expand in the lateral direction, Rayleigh-Taylor (RT) instability develops on the windward surface near the periphery, turning the drop from a bowl to a shape similar to a Sombrero hat ($t^*=1.4$ to 1.7). The drop at $t^*=1.7$ consists of both a backward bag near the central axis and a forward ring bag. This complex shape is the outcome of both the KH and RT instabilities and  this drop morphology has been observed for moderate We when $\eta$ is not too small \cite{marcotte_density_2019, jain_secondary_2019}. The high-speed gas blows in the forward ring bag and causes it to inflate rapidly, resulting in a fast decrease in the sheet thickness. At around $t^*=1.8$, holes appear in the forward ring bag. The expansion and merging of multiple  holes disintegrate the ring bag, forming a large number of small children drops, an unbroken backward bag, and a circular rim ($t^*=1.8$ to 2.0). To simulate the subsequent breakup of the remaining backward bag and rim, a longer simulation in a larger domain is required. Yet such a simulation is out of the scope of this paper since the purpose of the present test is to demonstrate the capability of the present methods. 

%%%%%%%%%%%%%% Figure 15 %%%%%%%%%%%%%%%%%%%

%%%%%%%%%%%%%% Figure 16 %%%%%%%%%%%%%%%%%%%

The vaporization here is driven by the superheated free stream, and the rate of vaporization depends on the vapor temperature gradient near the interface. It can be observed from Fig.~\ref{fig:acetone_drop_u_T} that the temperature gradient magnitude is higher on the windward side of the drop, where stronger vaporization occurs. The vapor generated is at a lower temperature ($T_\text{sat}$) compared to the free-stream temperature ($T_\infty$). {The low-temperature vapor is entrained in the wake and mixed with the free-stream hot vapor. As a result, the temperature gradient on the leeward side of the drop is lower and less vapor is produced, see Fig.~\ref{fig:acetone_drop_u_T}(c)-(d).}

\begin{figure}[htp]
	\begin{center}
		\includegraphics [width=1\columnwidth]{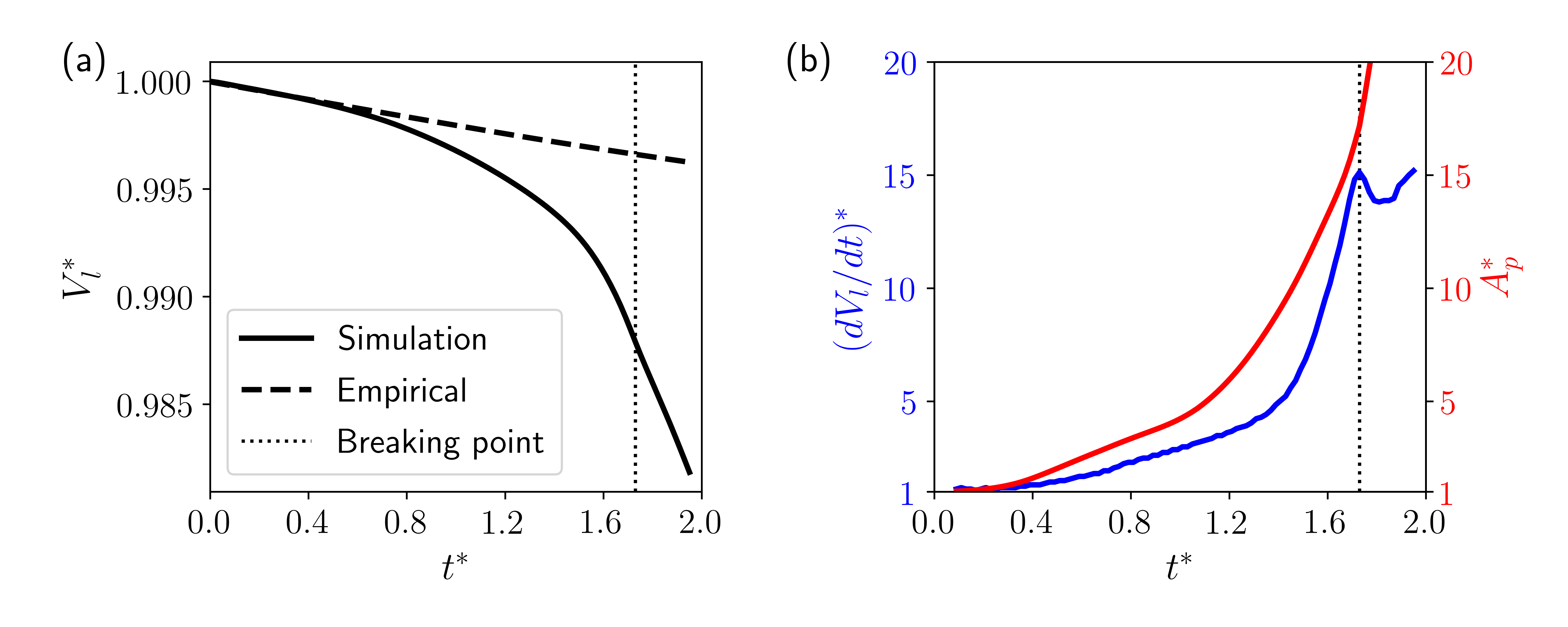}
	\end{center}
	\caption{Temporal evolution of (a) the drop volume ($V_l^*=V_l/V_{l,0} $) and (b) the rate of vaporization normalized by empirical vaporization rate ($[dV_l/dt]^*$) and the projected area of the droplet ($A_p^*=A_p/A_{p,0}$). The empirical correlation (Eq. \eqref{eq:empirical}) is also shown in (a) for comparison. The vertical dotted line indicates the onset of breakup of the forward ring bag ($t^*\approx1.7$). }
	\label{fig:fig_vol} 
\end{figure}

As the drop deforms, the drop surface area increases. Corresponding to that, the drop vaporization rate (the rate of loss of drop volume) also increases. The time evolution of the drop volume is shown in Fig.~\ref{fig:fig_vol}(a). The empirical relation for a non-deformable spherical drop (Eq. \eqref{eq:empirical}) is also plotted for comparison. When the drop is perfectly spherical, the volume decreases at a constant rate. For $t^*=0$ to $0.4$, the drop remains approximately spherical, see Fig.~\ref{fig:acetone_drop}. As a result, the drop volume decreases linearly in time, and the simulation results agree well with the empirical correlation, similar to the low-We drop considered previously in section~\ref{section:droplet}. As time evolves and the drop deformation becomes more significant, the vaporization rate goes up and  the drop volume decreases significantly faster, due to the increase of drop surface area, in particular the area of the windward surface. To better illustrate the relation between the drop vaporization rate and the surface area, Fig.~\ref{fig:fig_vol}(b) shows the temporal evolution of the normalized vaporization rate $(dV_l/dt)^*=(dV_l/dt)/(dV_l/dt)_0$ and the normalized projected frontal area of the drop, $A_p^*=A_p/A_{p,0}$. The results clearly show that the increase in frontal area is closely related to the increasing rate of volume loss. 

The inflation of the forward ring bag starts at approximately $t^*=1.6$, after which the increase of surface area seems to contribute little to the enhancement of vaporization, and $(dV_l/dt)^*$ reaches a plateau at about 15. This is probably due to the fact that the strong convective effect in the gas flow in the forward ring bag has reduced the temperature gradient magnitude near the interface, see Fig.~\ref{fig:acetone_drop_u_T}(e). After the forward ring bag breaks, the rapid vaporization of the small drops generated dominates the loss of the total liquid volume over time. Yet it should be noted that even though a high mesh resolution has been used in the present simulation, some of the tiny children droplets generated are still not well resolved. The temperature gradient and vaporization rate for those small drops may be underestimated. Since the Weber number for these tiny drops is very small, they typically exhibit a spherical shape with very mild deformation. Therefore, it will be more computationally efficient to model the vaporization of these tiny drops through the Lagrangian point-particle approach, instead of resolving them with the VOF method and a finer mesh, see for example the multiscale VOF-LPP model in the previous studies for drop dynamics \cite{herrmann_parallel_2010, ling_multiscale_2015, zuzio_improved_2018}. Nevertheless, such a model that can incorporate heat transfer and vaporization remains to be established. 

%%%%%%%%%%%%%% Figure 17 %%%%%%%%%%%%%%%%%%%

%===============================================================================
%  Conclusions
%===============================================================================
\section{Conclusions}
\label{conclusions}
A consistent volume-of-fluid method has been developed in the present study for the direct numerical simulation of interfacial multiphase flows with phase change. The geometric VOF method is based on Piece-wise Linear Interface Calculation (PLIC) reconstruction and the advection of momentum and energy is consistent with the VOF advection. A novel method is proposed to distribute the volumetric source, induced by  vaporization at the interfacial cells, to the neighboring pure gas cells. With this simple but elegant treatment, the velocity at the interfacial cells is not influenced by the Stefan vapor flow and can be used to advect the interface directly. The additional shifting of the interface toward the liquid side due to vaporization is handled geometrically, using the interface normal obtained in interface reconstruction. The present method does not require an additional pressure equation and projection as needed in previous studies. Furthermore, the Stefan flows near the interface do not exhibit any artificial numerical oscillations, as observed in previous studies, due to the estimate of the vaporization rate. 

The present methods have been implemented in the open-source solver \emph{Basilisk} using the octree/quad-tree mesh. The developed simulation framework has been tested against various benchmark cases. For all cases considered, the simulation results converge as the mesh is refined. The simulation results agree very well with the exact solutions for the 1D Stefan and sucking problems and the growth of a 3D spherical bubble in superheated liquid. The code has also been used to simulate a 2D film boiling problem and the simulation results agree well with previous numerical results obtained using a uniform mesh. 

To further test the capability of the present methods in resolving freely moving vaporizing drop, 2D axisymmetric simulations have been conducted to study the vaporization of a water drop in a high-temperature superheated vapor stream. The Weber number is 1.5 and thus the drop deformation is mild. A parametric study is carried out by varying the Reynolds number (Re) from 20 to 200. The simulation results agree very well with the empirical correlation for a spherical drop for all Re considered. The test results also affirm that the Stefan flow is important to the heat transfer between the drop and the surrounding vapor and also the vaporization rate. The drop volume loss due to vaporization will be overestimated if the Stefan flow is ignored. 

Finally, a fully 3D simulation was performed for the aerodynamic breakup of an acetone drop in a high-speed and high-temperature vapor stream. The free-stream properties are similar to the post-shock conditions for a planar shock with Mach number 1.6, though the compressibility effect is not considered. With the help of adaptive mesh refinement, we were able to accurately resolve the drop surface and the temperature field near the interface. The minimum cell size in the octree mesh is equivalent to 256 cells across the initial drop diameter. Due to the high Weber number ($\text{We}=62.7$), the drop breakup is in the multi-mode regime where the drop deforms to a Sombrero-hat shape, consisting of a backward bag at the center and a forward ring bag near the periphery. Eventually, the forward ring bag breaks, forming numerous small children droplets. 

The vaporization rate of the drop initially agrees with the empirical correlation, which is expected, since the drop is still approximately spherical. As the drop deformation becomes significant, the increase of frontal surface area results in a significantly increased rate of vaporization. The decrease of drop volume in time becomes nonlinear and is much faster than the empirical correlation for a spherical drop. The present results indicated that the effects of We and drop deformation are important to drop vaporization if the drop Weber number is finite, though a systematic parametric numerical investigation using the present simulation framework to fully characterize the effects of important parameters like We and Re will be relegated to our future work. 

%===============================================================================
%   Appendix
%===============================================================================

\appendix

%===============================================================================
%   Acknowledgments and bibliography
%=============================================================================== 
\section*{Acknowledgments}
This research was supported by the ACS Petroleum Research Fund (\#62481-ND9). BB has been supported by the Baylor University Postdoctoral Hiring Program, while YL has also been partially supported by the NSF grant (\#1942324). The authors also acknowledge the Extreme Science and Engineering Discovery Environment (XSEDE) and the Texas Advanced Computing Center (TACC) programs for providing the computational resources that have contributed to the research results reported in this paper. The Baylor High Performance and Research Computing Services (HPRCS) have been used to process the simulation data. The newly-developed methods have been implemented in the open-source multiphase flow solver \emph{Basilisk}, which is made available by St\'ephane Popinet and other collaborators.

\bibliography{references}

\begin{thebibliography}{10}
\expandafter\ifx\csname url\endcsname\relax
  \def\url#1{\texttt{#1}}\fi
\expandafter\ifx\csname urlprefix\endcsname\relax\def\urlprefix{URL }\fi
\expandafter\ifx\csname href\endcsname\relax
  \def\href#1#2{#2} \def\path#1{#1}\fi

\bibitem{lefebvre_atomization_2017}
A.~H. Lefebvre, V.~G. McDonell, Atomization and sprays, CRC press, 2017.

\bibitem{boyd_numerical_2019}
B.~Boyd, S.~Becker, Numerical modeling of the acoustically driven growth and
  collapse of a cavitation bubble near a wall, Physics of Fluids 31~(3) (2019)
  032102.

\bibitem{boyd_numerical_2018}
B.~Boyd, S.~Becker, Numerical modelling of an acoustically-driven bubble
  collapse near a solid boundary, Fluid Dynamics Research 50~(6) (2018) 065506.

\bibitem{boyd_beamed_2020}
B.~Boyd, S.~A. Suslov, S.~Becker, A.~D. Greentree, I.~S. Maksymov, Beamed {UV}
  sonoluminescence by aspherical air bubble collapse near liquid-metal
  microparticles, Scientific reports 10~(1) (2020) 1501.

\bibitem{tryggvason_direct_2011}
G.~Tryggvason, R.~Scardovelli, S.~Zaleski, Direct numerical simulations of
  gas-liquid multiphase flows, Cambridge University Press, 2011.

\bibitem{saurel_multiphase_1999}
R.~Saurel, R.~Abgrall, A multiphase {Godunov} method for compressible
  multifluid and multiphase flows, Journal of Computational Physics 150~(2)
  (1999) 425--467.

\bibitem{johnsen_implementation_2006}
E.~Johnsen, T.~Colonius, Implementation of {WENO} schemes in compressible
  multicomponent flow problems, Journal of Computational Physics 219~(2) (2006)
  715--732.

\bibitem{boyd_diffuse-interface_2021}
B.~Boyd, D.~Jarrahbashi, A diffuse-interface method for reducing spurious
  pressure oscillations in multicomponent transcritical flow simulations,
  Computers \& Fluids (2021) 104924.

\bibitem{boyd_numerical_2021}
B.~Boyd, D.~Jarrahbashi, Numerical study of the transcritical shock-droplet
  interaction, Physical Review Fluids 6~(11) (2021) 113601.

\bibitem{scardovelli_direct_1999}
R.~Scardovelli, S.~Zaleski, Direct numerical simulation of free-surface and
  interfacial flow, Annual Review of Fluid Mechanics 31~(1) (1999) 567--603.

\bibitem{unverdi_front-tracking_1992}
S.~O. Unverdi, G.~Tryggvason, A front-tracking method for viscous,
  incompressible, multi-fluid flows, Journal of Computational Physics 100
  (1992) 25--37.

\bibitem{sussman_level_1994}
M.~Sussman, P.~Smereka, S.~Osher, A level set approach for computing solutions
  to incompressible two-phase flow, Journal of Computational Physics 114 (1994)
  146--159.

\bibitem{sher_flash-boiling_2008}
E.~Sher, T.~Bar-Kohany, A.~Rashkovan, Flash-boiling atomization, Progress in
  Energy and Combustion Science 34 (2008) 417--439.

\bibitem{villegas_direct_2017}
L.~R. Villegas, S.~Tanguy, G.~Castanet, O.~Caballina, F.~Lemoine, Direct
  numerical simulation of the impact of a droplet onto a hot surface above the
  {Leidenfrost} temperature, International Journal of Heat and Mass Transfer
  104 (2017) 1090--1109.

\bibitem{son_temperature_2021}
J.~H. Son, I.~S. Park, Temperature changes around interface cells in a
  one-dimensional {Stefan} condensation problem using four well-known
  phase-change models, International Journal of Thermal Sciences 161 (2021)
  106718.

\bibitem{samkhaniani_numerical_2016}
N.~Samkhaniani, M.~Ansari, Numerical simulation of bubble condensation using
  {CF}-{VOF}, Progress in Nuclear Energy 89 (2016) 120--131.

\bibitem{sazhin_advanced_2006}
S.~S. Sazhin, Advanced models of fuel droplet heating and evaporation, Progress
  in Energy and Combustion Science 32~(2) (2006) 162--214.

\bibitem{shao_computational_2018}
C.~Shao, K.~Luo, M.~Chai, H.~Wang, J.~Fan, A computational framework for
  interface-resolved {DNS} of simultaneous atomization, evaporation and
  combustion, Journal of Computational Physics 371 (2018) 751--778.

\bibitem{irfan_front_2017}
M.~Irfan, M.~Muradoglu, A front tracking method for direct numerical simulation
  of evaporation process in a multiphase system, Journal of Computational
  Physics 337 (2017) 132--153.

\bibitem{safari_extended_2013}
H.~Safari, M.~H. Rahimian, M.~Krafczyk, Extended lattice {Boltzmann} method for
  numerical simulation of thermal phase change in two-phase fluid flow,
  Physical Review E 88~(1) (2013) 013304.

\bibitem{haghani-hassan-abadi_phase-change_2021}
R.~Haghani-Hassan-Abadi, A.~Fakhari, M.-H. Rahimian, Phase-change modeling
  based on a novel conservative phase-field method, Journal of Computational
  Physics 432 (2021) 110111.

\bibitem{gibou_level_2007}
F.~Gibou, L.~Chen, D.~Nguyen, S.~Banerjee, A level set based sharp interface
  method for the multiphase incompressible {Navier}–{Stokes} equations with
  phase change, Journal of Computational Physics 222~(2) (2007) 536--555.

\bibitem{lee_direct_2017}
M.~S. Lee, A.~Riaz, V.~Aute, Direct numerical simulation of incompressible
  multiphase flow with phase change, Journal of Computational Physics 344
  (2017) 381--418.

\bibitem{welch_volume_2000}
S.~W. Welch, J.~Wilson, A volume of fluid based method for fluid flows with
  phase change, Journal of Computational Physics 160~(2) (2000) 662--682.

\bibitem{hardt_evaporation_2008}
S.~Hardt, F.~Wondra, Evaporation model for interfacial flows based on a
  continuum-field representation of the source terms, Journal of Computational
  Physics 227~(11) (2008) 5871--5895.

\bibitem{ma_numerical_2013}
C.~Ma, D.~Bothe, Numerical modeling of thermocapillary two-phase flows with
  evaporation using a two-scalar approach for heat transfer, Journal of
  Computational Physics 233 (2013) 552--573.

\bibitem{sato_sharp-interface_2013}
Y.~Sato, B.~Ničeno, A sharp-interface phase change model for a
  mass-conservative interface tracking method, Journal of Computational Physics
  249 (2013) 127--161.

\bibitem{perez-raya_modeling_2016}
I.~Perez-Raya, S.~G. Kandlikar, Modeling of evaporation phenomenon considering
  liquid and vapor phase conduction effects: {Stefan} problems, in: {ASME} 2016
  14th {International} {Conference} on {Nanochannels}, {Microchannels}, and
  {Minichannels}, Vol.~1, American Society of Mechanical Engineers, Washington,
  DC, USA, 2016, p.~1.

\bibitem{datta_modeling_2017}
P.~Datta, A.~Chakravarty, K.~Ghosh, A.~Mukhopadhyay, S.~Sen, Modeling aspects
  of vapor bubble condensation in subcooled liquid using the {VOF} approach,
  Numerical Heat Transfer, Part A: Applications 72~(3) (2017) 236--254.

\bibitem{ding_volume_2017}
S.-T. Ding, B.~Luo, G.~Li, A volume of fluid based method for vapor-liquid
  phase change simulation with numerical oscillation suppression, International
  Journal of Heat and Mass Transfer 110 (2017) 348--359.

\bibitem{wilson_phase-change_2019}
J.~A. Wilson, M.~Haghshenas, R.~Kumar, Phase-change mechanism for evaporation
  in porous media using volume of fluid: {Implicit} formulation of interfacial
  temperature, International Communications in Heat and Mass Transfer 103
  (2019) 90--99.

\bibitem{wang_vaporization_2019}
Y.~Wang, V.~Yang, Vaporization of liquid droplet with large deformation and
  high mass transfer rate, {I}: {Constant}-density, constant-property case,
  Journal of Computational Physics 392 (2019) 56--70.

\bibitem{palmore_volume_2019}
J.~Palmore, O.~Desjardins, A volume of fluid framework for interface-resolved
  simulations of vaporizing liquid-gas flows, Journal of Computational Physics
  399 (2019) 108954.

\bibitem{scapin_volume--fluid_2020}
N.~Scapin, P.~Costa, L.~Brandt, A volume-of-fluid method for interface-resolved
  simulations of phase-changing two-fluid flows, Journal of Computational
  Physics 407 (2020) 109251.

\bibitem{bures_direct_2021}
L.~Bureš, Y.~Sato, Direct numerical simulation of evaporation and condensation
  with the geometric {VOF} method and a sharp-interface phase-change model,
  International Journal of Heat and Mass Transfer 173 (2021) 121233.

\bibitem{malan_geometric_2021}
L.~Malan, A.~Malan, S.~Zaleski, P.~Rousseau, A geometric {VOF} method for
  interface resolved phase change and conservative thermal energy advection,
  Journal of Computational Physics 426 (2021) 109920.

\bibitem{kharangate_review_2017}
C.~R. Kharangate, I.~Mudawar, Review of computational studies on boiling and
  condensation, International Journal of Heat and Mass Transfer 108 (2017)
  1164--1196.

\bibitem{tryggvason_direct_2005}
G.~Tryggvason, A.~Esmaeeli, N.~Al-Rawahi, Direct numerical simulations of flows
  with phase change, Computers \& Structures 83~(6-7) (2005) 445--453.

\bibitem{gao_effect_2022}
X.~Gao, J.~Chen, Y.~Qiu, Y.~Ding, J.~Xie, Effect of phase change on jet
  atomization: a direct numerical simulation study, Journal of Fluid Mechanics
  935 (2022) A16.

\bibitem{zhao_boiling_2022}
S.~Zhao, J.~Zhang, M.-J. Ni, Boiling and evaporation model for liquid-gas
  flows: {A} sharp and conservative method based on the geometrical {VOF}
  approach, Journal of Computational Physics 452 (2022) 110908.

\bibitem{johansen_cartesian_1998}
H.~Johansen, P.~Colella, A cartesian grid embedded boundary method for
  poisson’s equation on irregular domains, Journal of Computational Physics
  147~(1) (1998) 60--85.

\bibitem{schwartz_cartesian_2006}
P.~Schwartz, M.~Barad, P.~Colella, T.~Ligocki, A {Cartesian} grid embedded
  boundary method for the heat equation and {Poisson}’s equation in three
  dimensions, Journal of Computational Physics 211~(2) (2006) 531--550.

\bibitem{hsiang_drop_1995}
L.-P. Hsiang, G.~M. Faeth, Drop deformation and breakup due to shock wave and
  steady disturbances, International Journal of Multiphase Flow 21 (1995)
  545--560.

\bibitem{theofanous_physics_2008}
T.~G. Theofanous, G.~J. Li, On the physics of aerobreakup, Physics of Fluids
  20~(5) (2008) 052103.

\bibitem{meng_numerical_2018}
J.~C. Meng, T.~Colonius, Numerical simulation of the aerobreakup of a water
  droplet, Journal of Fluid Mechanics 835 (2018) 1108.

\bibitem{jain_secondary_2019}
S.~S. Jain, N.~Tyagi, R.~S. Prakash, R.~Ravikrishna, G.~Tomar, Secondary
  breakup of drops at moderate {Weber} numbers: {Effect} of {Density} ratio and
  {Reynolds} number, International Journal of Multiphase Flow 117 (2019)
  25--41.

\bibitem{jackiw_aerodynamic_2021}
I.~M. Jackiw, N.~Ashgriz, On aerodynamic droplet breakup, Journal of Fluid
  Mechanics 913 (2021) A33.

\bibitem{duke-walker_evaporation_2021}
V.~Duke-Walker, W.~C. Maxon, S.~R. Almuhna, J.~A. McFarland, Evaporation and
  breakup effects in the shock-driven multiphase instability, Journal of Fluid
  Mechanics 908 (2021) A13.

\bibitem{dahal_numerical_2017}
J.~Dahal, J.~A. McFarland, A numerical method for shock driven multiphase flow
  with evaporating particles, Journal of Computational Physics 344 (2017)
  210--233.

\bibitem{gallot-lavallee_large_2021}
S.~Gallot-Lavallée, W.~P. Jones, A.~J. Marquis, Large eddy simulation of an
  ethanol spray flame with secondary droplet breakup, Flow, Turbulence and
  Combustion 107~(3) (2021) 709--743.

\bibitem{salman_lagrangian_2004}
H.~Salman, M.~Soteriou, Lagrangian simulation of evaporating droplet sprays,
  Physics of Fluids 16~(12) (2004) 4601--4622.

\bibitem{maxey_equation_1983}
M.~R. Maxey, J.~J. Riley, Equation of motion for a small rigid sphere in a
  nonuniform flow, Physics of Fluids 26 (1983) 883--889.

\bibitem{balachandar_turbulent_2010}
S.~Balachandar, J.~K. Eaton, Turbulent dispersed multiphase flow, Annual Review
  of Fluid Mechanics 42 (2010) 111--133.

\bibitem{renksizbulut_experimental_1983}
M.~Renksizbulut, M.~C. Yuen, Experimental study of droplet evaporation in a
  high-temperature air stream, Journal of Heat Transfer 105~(2) (1983)
  384--388.

\bibitem{popinet_gerris_2003}
S.~Popinet, Gerris: a tree-based adaptive solver for the incompressible {Euler}
  equations in complex geometries, Journal of Computational Physics 190~(2)
  (2003) 572--600.

\bibitem{popinet_accurate_2009}
S.~Popinet, An accurate adaptive solver for surface-tension-driven interfacial
  flows, Journal of Computational Physics 228~(16) (2009) 5838--5866.

\bibitem{popinet_quadtree-adaptive_2015}
S.~Popinet, A quadtree-adaptive multigrid solver for the
  {Serre}–{Green}–{Naghdi} equations, Journal of Computational Physics 302
  (2015) 336--358.

\bibitem{zhang_modeling_2020}
B.~Zhang, S.~Popinet, Y.~Ling, Modeling and detailed numerical simulation of
  the primary breakup of a gasoline surrogate jet under non-evaporative
  operating conditions, International Journal of Multiphase Flow 130 (2020)
  103362.

\bibitem{zhang_direct_2021}
B.~Zhang, B.~Boyd, Y.~Ling, Direct numerical simulation of compressible
  interfacial multiphase flows using a mass-momentum-energy consistent
  volume-of-fluid method, Computers \& Fluids (2021) 105267.

\bibitem{georgoulas_enhanced_2017}
A.~Georgoulas, M.~Andredaki, M.~Marengo, An enhanced {VOF} method coupled with
  heat transfer and phase change to characterise bubble detachment in saturated
  pool boiling, Energies 10~(3) (2017) 272.

\bibitem{weymouth_conservative_2010}
G.~Weymouth, D.~K.-P. Yue, Conservative {Volume}-of-{Fluid} method for
  free-surface simulations on {Cartesian}-grids, Journal of Computational
  Physics 229~(8) (2010) 2853--2865.

\bibitem{fuster_all-mach_2018}
D.~Fuster, S.~Popinet, An all-{Mach} method for the simulation of bubble
  dynamics problems in the presence of surface tension, Journal of
  Computational Physics 374 (2018) 752--768.

\bibitem{francois_balanced-force_2006}
M.~M. Francois, S.~J. Cummins, E.~D. Dendy, D.~B. Kothe, J.~M. Sicilian, M.~W.
  Williams, A balanced-force algorithm for continuous and sharp interfacial
  surface tension models within a volume tracking framework, Journal of
  Computational Physics 213 (2006) 141--173.

\bibitem{van_hooft_towards_2018}
J.~A. van Hooft, S.~Popinet, C.~C. van Heerwaarden, S.~J.~A. van~der Linden,
  S.~R. de~Roode, B.~J.~H. van~de Wiel, Towards adaptive grids for atmospheric
  boundary-layer simulations, Boundary-Layer Meteorology 167~(3) (2018)
  421--443.

\bibitem{aulisa_interface_2007}
E.~Aulisa, S.~Manservisi, R.~Scardovelli, S.~Zaleski, Interface reconstruction
  with least-squares fit and split advection in three-dimensional {Cartesian}
  geometry, Journal of Computational Physics 225~(2) (2007) 2301--2319.

\bibitem{vaudor_consistent_2017}
G.~Vaudor, T.~Ménard, W.~Aniszewski, M.~Doring, A.~Berlemont, A consistent
  mass and momentum flux computation method for two phase flows. {Application}
  to atomization process, Computers \& Fluids 152 (2017) 204--216.

\bibitem{arrufat_momentum-conserving_2020}
T.~Arrufat, M.~Crialesi-Esposito, D.~Fuster, Y.~Ling, L.~Malan, S.~Pal,
  R.~Scardovelli, G.~Tryggvason, S.~Zaleski, A momentum-conserving, consistent,
  {Volume}-of-{Fluid} method for incompressible flow on staggered grids,
  Computers \& Fluids 215 (2020) 104785.

\bibitem{lopez-herrera_electrokinetic_2015}
J.~López-Herrera, A.~Gañán-Calvo, S.~Popinet, M.~Herrada, Electrokinetic
  effects in the breakup of electrified jets: {A} {Volume}-{Of}-{Fluid}
  numerical study, International Journal of Multiphase Flow 71 (2015) 14--22.

\bibitem{bell_second-order_1989}
J.~B. Bell, P.~Colella, H.~M. Glaz, A second-order projection method for the
  incompressible navier-stokes equations, Journal of Computational Physics
  85~(2) (1989) 257--283.

\bibitem{lalanne_numerical_2021}
C.~Lalanne, Q.~Magdelaine, F.~Lequien, J.-M. Fullana, Numerical model using a
  {Volume}-{Of}-{Fluid} method for the study of evaporating sessile droplets in
  both unpinned and pinned modes, European Journal of Mechanics - B/Fluids 89
  (2021) 267--273.

\bibitem{berenson_film-boiling_1961}
P.~J. Berenson, Film-boiling heat transfer from a horizontal surface, Journal
  of Heat Transfer 83~(3) (1961) 351--356.

\bibitem{klimenko_film_1981}
V.~Klimenko, Film boiling on a horizontal plate — new correlation,
  International Journal of Heat and Mass Transfer 24~(1) (1981) 69--79.

\bibitem{sun_modeling_2014}
D.~Sun, J.~Xu, Q.~Chen, Modeling of the evaporation and condensation
  phase-change problems with {FLUENT}, Numerical Heat Transfer, Part B:
  Fundamentals 66~(4) (2014) 326--342.

\bibitem{esmaeeli_computations_2004}
A.~Esmaeeli, G.~Tryggvason, Computations of film boiling. {Part} {I}: numerical
  method, International Journal of Heat and Mass Transfer 47~(25) (2004)
  5451--5461.

\bibitem{tomar_numerical_2005}
G.~Tomar, G.~Biswas, A.~Sharma, A.~Agrawal, Numerical simulation of bubble
  growth in film boiling using a coupled level-set and volume-of-fluid method,
  Physics of Fluids 17~(11) (2005) 112103.

\bibitem{pilch_use_1987}
M.~Pilch, C.~A. Erdman, Use of breakup time data and velocity history data to
  predict the maximum size of stable fragments for acceleration-induced breakup
  of a liquid drop, International Journal of Multiphase Flow 13 (1987)
  741--757.

\bibitem{hsiang_near-limit_1992}
L.-P. Hsiang, G.~M. Faeth, Near-limit drop deformation and secondary breakup,
  International Journal of Multiphase Flow 18 (1992) 635--652.

\bibitem{joseph_breakup_1999}
D.~D. Joseph, J.~Belanger, G.~S. Beavers, Breakup of a liquid drop suddenly
  exposed to a high-speed airstream, International Journal of Multiphase Flow
  25~(6) (1999) 1263--1303.

\bibitem{guildenbecher_secondary_2009}
D.~R. Guildenbecher, C.~López-Rivera, P.~E. Sojka, Secondary atomization,
  Experiments in Fluids 46~(3) (2009) 371.

\bibitem{theofanous_physics_2012}
T.~G. Theofanous, V.~V. Mitkin, C.~L. Ng, C.~H. Chang, X.~Deng, S.~Sushchikh,
  The physics of aerobreakup. {II}. {Viscous} liquids, Physics of Fluids 24
  (2012) 022104.

\bibitem{mahmood_effects_2021}
T.~H. Mahmood, Y.~Ling, Effects of {Reynolds} number on aerobreakup of viscous
  drops, in: Proceedings of {ICLASS} 2021: 15th {Triennial} {International}
  {Conference} on {Liquid} {Atomization} and {Spray} {Systems}, Vol.~1, 2021,
  p.~1.

\bibitem{balachandar_scaling_2009}
S.~Balachandar, A scaling analysis for point particle approaches to turbulent
  multiphase flows, International Journal of Multiphase Flow 35 (2009)
  801--810.

\bibitem{wert_rationally-based_1995}
K.~L. Wert, A rationally-based correlation of mean fragment size for drop
  secondary breakup, International Journal of Multiphase Flow 21 (1995)
  1063--1071.

\bibitem{dai_temporal_2001}
Z.~Dai, G.~M. Faeth, Temporal properties of secondary drop breakup in the
  multimode breakup regime, International Journal of Multiphase Flow 27 (2001)
  217--236.

\bibitem{kuo_maximum_2022}
C.-W. Kuo, M.~F. Trujillo, A maximum entropy formalism model for the breakup of
  a droplet, Physics of Fluids 34~(1) (2022) 013315.

\bibitem{ranger_aerodynamic_1969}
A.~A. Ranger, J.~A. Nicholls, Aerodynamic shattering of liquid drops., AIAA
  Journal 7 (1969) 285--290.

\bibitem{haywood_detailed_1989}
R.~J. Haywood, R.~Nafziger, M.~Renksizbulut, A detailed examination of gas and
  liquid phase transient processes in convective droplet evaporation, Journal
  of Heat Transfer 111~(2) (1989) 495--502.

\bibitem{chiang_numerical_1992}
C.~H. Chiang, M.~S. Raju, W.~A. Sirignano, Numerical analysis of convecting,
  vaporizing fuel droplet with variable properties, International Journal of
  Heat and Mass Transfer (1992) 18.

\bibitem{yuen_heat-transfer_1978}
M.~Yuen, L.~Chen, Heat-transfer measurements of evaporating liquid droplets,
  International Journal of Heat and Mass Transfer 21~(5) (1978) 537--542.

\bibitem{renksizbulut_numerical_1983}
M.~Renksizbulut, M.~C. Yuen, Numerical study of droplet evaporation in a
  high-temperature stream, Journal of Heat Transfer 105~(2) (1983) 389--397.

\bibitem{marcotte_density_2019}
F.~Marcotte, S.~Zaleski, Density contrast matters for drop fragmentation
  thresholds at low {Ohnesorge} number, Physical Review Fluids 4~(10) (2019)
  103604.

\bibitem{herrmann_parallel_2010}
M.~Herrmann, A parallel {Eulerian} interface tracking/{Lagrangian} point
  particle multi-scale coupling procedure, Journal of Computational Physics 229
  (2010) 745--759.

\bibitem{ling_multiscale_2015}
Y.~Ling, S.~Zaleski, R.~Scardovelli, Multiscale simulation of atomization with
  small droplets represented by a {Lagrangian} point-particle model,
  International Journal of Multiphase Flow 76 (2015) 122--143.

\bibitem{zuzio_improved_2018}
D.~Zuzio, J.-L. Estivalezes, B.~DiPierro, An improved multiscale
  {Eulerian}-{Lagrangian} method for simulation of atomization process,
  Computers \& Fluids 176 (2018) 285--301.

\end{thebibliography}

\end{document}